\documentclass[aps,amsfonts,showpacs,nofootinbib,
eqsecnum,floats,amssymb,amsmath]{revtex4}
\usepackage{graphicx}
\usepackage{amsmath}
\usepackage{amsfonts}
\usepackage{amssymb}
\def\@ApndToks#1#2{\edef\@act{\noexpand#1={\the#1#2}}\@act}

\def\@begineqtableau#1#2{
  \vcenter\bgroup\openup1\jot
    \mathsurround=0pt  \everymath={\displaystyle}
    \dimen0=#1  \count0=#2  \toks0={\strut}  \toks1={##}
        \def\\{\crcr\noalign{\vskip\dimen0}}
    \ifnum\count0 > 0
      \loop  \advance\count0 by -1
        \@ApndToks{\toks0}{$\hfil\the\toks1$&${}\the\toks1\hfil$}
      \ifnum\count0 > 0
        \@ApndToks{\toks0}{&}
      \repeat
    \else  \@ApndToks{\toks0}{\hfil$\the\toks1$\hfil}  \fi
    \edef\@act{\noexpand\ialign\bgroup\the\toks0\noexpand\crcr}  \@act}
\def\@endeqtableau{\crcr\egroup\egroup}
\makeatother
\def\beq{\begin{eqnarray}}
\def\eeq{\end{eqnarray}}
\def\be{\begin{equation}}
\def\ee{\end{equation}}

\def\={\triangleq}
\def\nn{\nonumber\\}

\def\ub#1{\underleftarrow{#1}}

\def\mbf#1{\mbox{\boldmath${#1}$}}

\def\grad{\nabla}
\def\lie{\pounds}
\def\l{\ell}

\def\bm{\bar m}
\def\bu{\bar U}
\def\bepsilon{\bar\epsilon}
\def\balpha{\bar\alpha}
\def\bbeta{\bar\beta}
\def\bpi{\bar\pi}
\def\bmu{\bar\mu}
\def\blambda{\bar\lambda}
\def\bdelta{\bar\delta}
\def\bfA{{\bf A}}
\def\bfF{{\bf F}}
\def\bfT{{\bf T}}

\def\sbfF{{{}^{\ast}\bf F}}

\begin{document}
\title{Laws of Black Hole Mechanics from Holst Action }

\author{Ayan Chatterjee}\email{ayan.chatterjee@saha.ac.in}
\author{Amit Ghosh}\email{amit.ghosh@saha.ac.in}

\affiliation{Theory Division, Saha Institute of
Nuclear Physics, Kolkata 700064, India}

\begin{abstract}
The formulation of Weak Isolated Horizons (WIH)
based on the Isolated Horizon formulation of black hole horizons
is reconsidered. The  first part of the 
paper deals with the derivation of laws
of mechanics of a WIH. While the zeroth law follows from
the WIH boundary conditions, first law depends on the action chosen.
We construct the covariant phase space
for a spacetime having an WIH as inner boundary for the Holst action.
This requires the introduction of new potential functions
so that the symplectic structure is
foliation independent. We show that a precise cancellation
among various terms leads to the usual first law for WIH.
Subsequently, we show from the same covariant phase space that for spherical horizons, the
topological theory on the inner boundary is a $U(1)$
Chern-Simons theory.
\end{abstract}

\pacs{04070B, 0420}

\maketitle

\section{Introduction} 

In general relativity black holes are exact solutions of Einstein's
equations and are the {\em simplest macroscopic objects} of Nature
\cite{Chandra}. Therefore, it is only natural that such objects are
perfect laboratories for the search of the quantum theory of general
relativity (much like the way solitons are ideal laboratories of any
non-perturbative quantum theory of non-abelian gauge fields
interacting with scalar fields). In the last century one of the major
development in the study of black hole physics came through the
realization that these solutions are analogous to some macrostates in
thermal equilibrium. Dynamical processes involving only black holes
obey laws that are qualitatively similar to the four laws of
thermodynamics - the zeroth, first, second and the third laws. The
proofs of these laws, called the laws of black hole dynamics, were
first given by \cite{bch}, where black hole spacetimes were supposed
to contain some event horizons. However, it was soon  realized that
the notion of event horizons is too impractical. The global features
of such horizons (for example one needs the entire asymptotic future
null infinity to know whether an event horizon is present) percolate
everywhere in the derivation of the said four laws of black hole
dynamics \cite{bch}, making the laws too abstract and unsuitable for
use in many practical situations.  
notion of event horizons (see \cite{abf} for details).  Killing
horizons, which were introduced as a practical alternative to the
event horizons, are closer to reality. First of all, Killing horizons
are almost local, requiring the Killing vector only on and in the
vicinity of the black hole horizon. The laws of black hole mechanics
were also proved in this setting \cite{Wald, Wald_Racz}, although it had
difficulties in handling the extremal black holes at the same footing
of the non-extremal ones. One difficulty being that the
proofs for the laws of Killing horizon dynamics crucially depends on
the existence of a bifurcating two-sphere at the horizon, which are
absent for the extremal black holes. The precise dependence
on the bifurcation two-sphere goes as follows: For non-extremal horizons the zeroth law
states that the the surface gravity is non-zero and constant
on the Killing horizon. Such a law holds if and only if the horizon is extendible to
a bifurcating horizon. Furthermore, the Noether charge coming as a surface integral 
over a cross-section of the horizon, which is defined to be the \emph{entropy}/$2\pi$
in this case, is expressed in terms of all dynamical fields, their derivatives and
also on the Killing field $\chi^a$ and its derivatives. However, the
explicit dependence on  $\chi^a$ can be eliminated provided one
uses the cross-section to be the bifurcation two-sphere.  One can also eliminate the second and
the higher derivatives of $\chi^a$ by using Killing vector identities,
leaving the entropy as a function of $\chi^a$ and  $\nabla_{a}\chi^b$.
The contribution of the term linear in
$\chi^a$ vanishes at the bifurcation point since the Killing vector
$\chi^a$ vanishes at that point.  Finally, at the bifurcation surface one has
$\nabla_{a}\chi_b=\epsilon_{ab}$, where $\epsilon_{ab}$ denotes the
binormal.  Thus, at the bifurcation point all explicit reference to
the Killing field can be eliminated from the Noether charge $Q$.  Thus $Q$ evaluated at
the bifurcation two-sphere defines the entropy only as a function of local geometric
quantities such as the metric, the matter fields and their derivatives
\cite{Wald}. Using the extension of the horizon one can then evaluate the entropy as an
integral over an arbitrary section of the Killing horizon rather than
on the bifurcation two-sphere alone, provided  the surface gravity is
\emph{constant and nonvanishing} somewhere on the Killing horizon.

Isolated horizon \cite{abf_letter,abf,afk,ak_lr} is another local
and practical alternative to a black hole horizon whose  descriptions
require the existence of some marginally trapped surfaces at the
horizon. Since there is no explicit reference to any Killing vector,
isolated horizons claim to do a better job than the Killing ones (see
\cite{ak_lr, Wald_lr, G_J_pr} for a detailed comparison and the recent
surveys). In fact, a large number of black hole horizons are isolated
horizons but not Killing horizons. The reason for this enhancement in
the space of solutions is that the boundary conditions defining an
isolated horizon are weaker compared to the ones defining a Killing
horizon. For example, as a consequence of the isolated horizon
boundary conditions it is seen that such horizons admit a Killing
vector field only on the horizon, whereas a Killing horizon requires
such a Killing vector field in some neighborhood of the horizon. The
isolated horizon formalism was initially formulated in terms of canonical variables
leading to the canonical phase space \cite{abf_letter,abf}. This
formulation also used to show that the effective field theory on the
isolated horizon is  a Chern-Simons theory \cite{ack}. The detailed
calculation, extensions and other consequences are discussed in \cite{abk}
This also led to the calculation of the entropy done first in
\cite{abck}.
The covariant
formulation was discussed in \cite{afk}. The various extensions of the
isolated horizons and its ramifications were discussed in a series of
papers \cite{abl}.

One important advantage of the isolated horizons over the Killing
ones is that it can deal with the extremal solutions at the same
footing of non-extremal ones \cite{cg}.  This requires the standard
isolated horizon boundary conditions to be weakened enough so as to
contain the extremal and non-extremal horizons as part of the same
phase space.   The standard
formulation of isolated horizons takes a rigid class of null normals
(which are Killing vectors only on the horizon) where the null vector
fields associated with a null surface are allowed to rescale only by a
positive constant. While this is definitely a possibility, the isolated
horizons may actually admit a much larger class of null vectors. The
new formulation, called {\em weak isolated horizons} (WIH), proposed
in \cite{cg}, relaxes this rigidness and allows rescaling by a class
of functions.  This new rescaling opens up the possibility of
extending the space of solution of an isolated horizon. The standard
formulation of isolated horizons place extremal and non-extremal
solutions in two distinct phase spaces, much like the way they have
been treated in the Killing horizon formulations; whereas in WIH, one
gets a single unified space of solutions that contain both types of
solutions. This implies that the laws of mechanics of WIHs, with these
improved set of boundary conditions, encompass both extremal and
non-extremal solutions at one go.

In this paper, we shall derive the laws of mechanics of an WIH from
Holst's action \cite{Holst} from a totally covariant framework. In the
framework of Loop Quantum Gravity (LQG) this action is a natural
starting point than the Palatini action. From this action we construct
the phase space for a solution having a WIH as an inner boundary (this
means that the spacetime admits an inner boundary which, in the
present case, satisfies the WIH boundary conditions). The symplectic
structure for this phase space is obtained, from which we prove the
first law of black hole mechanics. From this covariant symplectic
structure one then finds that the effective theory at the spherically
symmetric WIH is precisely a $U(1)$ Chern-Simons theory. The Chern-
Simons one- form gauge field is such that it does not depend on the
extremal or non- extremal nature of the horizon. This shows that the
effective theory for the spherical symmetric horizons, extremal or
non- extremal, is a $U(1)$ Chern-Simons theory. Correspondingly, the
entropy of these horizons will again be proportional to area of the
horizon. That the effective topological theory
on the horizon is a $U(1)$ Chern-Simons theory  was
also shown in \cite{abck, abk}. However in the present calculation, we
carefully derive the laws of black hole mechanics from a completely
covariant formulation taking into account the weakest possible
boundary conditions for a black hole horizon and then reinforce the
claims that the surface symplectic structure of the WIH is that of a
Chern- Simons theory.

The plan of the paper is as follows. We first recall the boundary
conditions of a Weak Isolated Horizon (WIH). Then, we derive some key
consequences of these boundary conditions. For example, the boundary
conditions result in the zeroth law of black hole mechanics provided
we restrict the equivalence class of null normals on the WIH. These
boundary conditions are equally applicable to extremal as well as
non-extremal WIH. Then, using the Holst action we show that the
principle of least action is well defined in presence of some
appropriate boundary terms. In the next step we construct the
symplectic structure in the space of solutions in which each solution
contains a WIH as its inner boundary. We then derive the 1st law of
black hole mechanics using this symplectic structure. We also derive
the Chern-Simons symplectic structure on the horizon from this
formulation.

\section{Weak Isolated Horizons}

We now give an introduction to the idea of weak isolated horizons \cite{cg}.
Let us consider $\cal M$ to be a four-manifold 
equipped with a metric $g_{ab}$ of signature
$(-,+,+,+)$. Our notations and conventions closely follow that of \cite{cg, afk}.
${\Delta}$ is a null hypersurface in $\cal M$ of which $\ell^a$ is a future
directed null normal. However, if $\ell^a$ is a null normal, so is
$\xi\ell^a$, where $\xi$ is any arbitrary positive function on $\Delta$. Thus, $\Delta$
naturally admits an equivalence class of null normals $[\,\xi\ell^a\,]$. We
denote by $q_{ab}\triangleq g_{\ub{ab}}$ the degenerate intrinsic metric on
$\Delta$ induced by $g_{ab}$ (indices that are not explicitly intrinsic on
$\Delta$ will be pulled back and $\triangleq$ means that the equality holds
{\em only on} $\Delta$). The tensor $q^{ab}$ will be an {\em inverse} of
$q_{ab}$ if it satisfies $q^{ab}q_{ac}q_{bd}\triangleq q_{cd}$. The expansion
$\theta_{(\ell\,)}$ of the null normal $\ell^a$ is then defined by
$\theta_{(\ell\,)}=q^{ab}\nabla_a\ell_b$, where $\nabla_a$ is the covariant
derivative compatible with $g_{ab}$.

The null surface $\Delta$ introduced above is an arbitrary 
null surface equipped with an equivalence class of null
normals $[\xi\l^a]$. 
The conditions on $\Delta$ are too general to make it resemble a
black hole horizon. To enrich $\Delta$ with
useful and interesting information, we need to impose
some restrictions on this surface.
The idea is that we endow a minimal set of conditions
on the null hypersurface $\Delta$ so that it behaves as a black hole
horizon. As we shall see, the zeroth law and the first law of black hole
mechanics will naturally follow from these conditions. These definitions
will be local and only provides a construction of black hole horizon and do not 
define a black hole spacetime which is a global object. However, if there is 
a global solution, like the Schwarzschild solution, then these conditions
will be satisfied.

\subsection{The First Set of Boundary Conditions}
We shall now introduce the set of boundary conditions to be imposed on the
null surface $\Delta$ so that effectively the surface behaves as a black hole
horizon. The boundary conditions that are proposed here are the 
least number of conditions that are necessary for a generic black hole
horizon. Since the null surface has an equivalence class
of null normals $[\xi\l^a]$ as its generators, it is natural
to impose the boundary conditions on all of these null normals \emph{i.e.} the boundary conditions 
has to hold for the entire equivalence class $[\xi\ell^a]$.

The null surface $\Delta$ generated by the equivalence class $[\xi\ell^a]$ will be called a \textit{non-expanding
horizon} (NEH) in $({\cal M},g_{ab})$ if the following conditions are satisfied \cite{cg}:
\begin{enumerate}
\item $\Delta$ is topologically $S^2 \times \mathbb{R}$.
\item The expansion $\theta_{(\xi\ell)}\=0$ for any $~\xi\ell^a$ in the
  equivalence class.
\item The equations of motion and energy conditions
 hold on the surface $\Delta$ and the vector field $-T^{a}_{b}\xi\ell^b$
is future directed and causal. 
\end{enumerate}
There are some important points to note in the boundary conditions presented here. 
Firstly,  all boundary conditions are intrinsic
to $\Delta$. This implies that to describe NEH, one needs no reference to the spacetime in the exterior.
Also, the definition doe not involve Killing vectors although, as we shall see,
the boundary conditions imply the presence of a Killing vector on the null
surface $\Delta$. Of these boundary conditions, the first one 
is just a topological restriction on the horizon and has no reference to the 
equivalence class of null normals. The second boundary condition is the
most important of the conditions. The expansion freeness is a special
requirement for any isolated black hole horizon \emph {i.e} it holds only for those 
null surfaces which are black holes. Any null surface will not satisfy this condition.
For example, the Minkowski light cone
does not satisfy the expansion free condition. This is because the Minkowski light cone
is not  a black hole horizon although it appears so for the Rindlar observers.
One might be tempted to infer that the second condition implies infinite number
of boundary conditions to be imposed on each of the infinite number of null 
normals in the equivalence class $[\xi\ell^a]$. This however is not true. In fact,
it is enough that the expansion corresponding to any one null normal is zero.
The quantities involved in the boundary conditions are such that once these conditions are satisfied by one
null vector $\ell^a$, then these are also obeyed by every null vector in the
class $[\,\xi\ell^a\,]$. This is true also for the  expansion-free condition,
since $\theta_{(\xi\ell\,)} \triangleq\xi\theta_{(\ell\,)}$. So, for all the
conditions defining a NEH, it is sufficient that these are satisfied by only
{\em one} normal vector field in the class $[\,\xi\ell^a\,]$
The third boundary condition ensures that the equation of motion of all
fields are satisfied on $\Delta$. This condition only allows those fields
which satisfy the dominant energy condition. This requirement also
holds true for all the null normals in the equivalence class $[\,\xi\ell^a\,]$ if
it holds true for one.

\subsection{Consequences of the Boundary Conditions} The above boundary
conditions have important consequences for the kinematical structure of the
horizon. First of all, note that one can have important simplifications
for the null surface $\Delta$ which are quite independent of the boundary conditions.
We note them below.
Since any $\l^a$ in $[\xi\l^a]$ is normal to $\Delta$, these are twist free. 
Next, because the surface is null, the normal vector is also the tangent vector.
The tangent vector fields are tangent to the generators of the surface. It can be shown that
these generators are geodesics or in other words,  each $\xi\ell^a$ in $[\xi\l^a]$ is   
geodetic, {\em i.e.}
\begin{equation}\label{eqgeodetic}
\xi\ell^a\grad_a(\xi\ell^b)\triangleq\kappa_{(\xi\ell\,)}\xi\ell^b
\end{equation}
where $\kappa_{(\xi\ell\,)}$ is the acceleration of $\xi\ell^a$. It can be easily
deduced from (\ref{eqgeodetic}) that the acceleration varies in the 
equivalence class 
\begin{equation}\label{kappa_eq_class}
\kappa_{(\xi\ell\,)}=\xi\kappa_{(\ell\,)}+\lie_\ell\,\xi
\end{equation}
We shall always work with the null tetrad basis $(\ell, n, m,
\bar{m})$ such that $1\!=\!-n\cdot\ell=\!m\cdot\bar m$ and all other scalar
products vanish. This is specially suited for the problem since one of the 
null normals $\l^a$ matches with one of the vectors in the equivalence
class $[\xi\l^a]$. The spacetime metric is then given by $g_{ab}=-2\l_{(a}
n_{b)}+ 2 m_{(a} \bar m_{b)}$.

Given the simplification that any null normal $\xi\ell^a$ is twist-free and the second boundary
condition that  any null normal $\xi\ell^a$ is expansion- free, the Raychaudhuri
equation becomes
\begin{equation}\label{eqray}
0\triangleq\lie_{\xi\ell}\,\theta_{(\xi\ell\,)}\triangleq-|\sigma_{(\xi\ell\,)}|^2
-\xi^2R_{ab}\ell^a\ell^b
\end{equation}
where, $\sigma_{(\xi\ell\,)}=m^am^b\nabla_a(\xi\ell_b)$ is the shear of
$\xi\ell^a$. By using the energy conditions and Einstein equations, it can be
shown that 
both terms on the right hand side of (\ref{eqray}) vanish independently on $\Delta$. Therefore
every null normal $\xi\ell^a$ in the equivalence class is also shear-free.
Again, note that if any one of the null normals $\l^a$ is shear free, all
the null normals in the equivalence class $[\xi\l^a]$ are shear free too. In short,
every null normal in the equivalence class $[\xi\l^a]$ is twist- free, shear- free
and expansion- free.
These conditions imply that there exists an one-form $\omega^{(\ell\,)}_{a}$ on $\Delta$,
depending on $\l^a$ such that 
\begin{equation}\label{omega_def}
\nabla_{\ub{a}}\l^b\=:\omega^{(\ell\,)}_{a}\l^b
\end{equation}
The one-form defined in (\ref{omega_def}) plays an important role in the whole analysis. It is 
also clear that since the one- form $\omega^{(\l)}_a$ depends on the null normals, it
varies in the class $[\,\xi\ell^a\,]$. The variation is 
\begin{equation}\label{omega_exp_eqclass}
{\omega}^{(\xi\ell\,)}= \omega^{(\ell\,)}+d\ln\xi
\end{equation}
where $d$ is the exterior derivative in $\Delta$. Since the pull-back of the
one-form $\xi\ell_a$ is zero on $\Delta$, it follows that every $\xi\ell^a$ 
in the class \emph{is a Killing vector on} $\Delta$, namely $\lie_{(\xi\ell\,)}\,q_{ab}\triangleq
0$. A straightforward calculation (using results of appendix (\ref{appb})) shows that the 
curvature of $\omega^{(\xi\ell\,)}$
\begin{equation} \label{curvomega}
d\omega^{(\xi\ell\,)}\triangleq 2({\rm Im}\Psi_2)\,{}^2\mbf{\epsilon}\;,
\end{equation}
where ${\rm Im}\Psi_2=C_{abcd}\ell^am^b\bar m^cn^d$ is a complex scalar, 
$C_{abcd}$ is the Weyl-tensor and ${}^2\mbf{\epsilon}=im\wedge\bar m$ is the 
area two- form on the cross- sections of $\Delta$. Again from equation
(\ref{omega_exp_eqclass}), since the
one form $\omega^{(\xi\ell\,)}$ varies like a $U(1)$ gauge field,
the equation (\ref{curvomega}) will hold true for all $\omega^{(\xi\ell\,)}$
corresponding to the vectors in the equivalence class.
The Killing equations imply that the area two- form ${}^2\mbf\epsilon$ 
of the cross-section is preserved under Lie-flow of every 
$\xi\ell^a$ in the class, $\lie_{(\xi\ell\,)}~{}^2\mbf{\epsilon}_{ab}\triangleq 0$.

\subsection{Weak Isolated Horizon and the Zeroth Law}
Let us recall that from the point of view of boundary conditions, all
the horizons generated by the null normals in the equivalence class
$[\xi\ell^a]$ are equivalent. In other words, the boundary 
conditions are oblivious to the equivalence class and cannot
prefer one horizon generated by say $\l^a$, over another generated
by $\xi\l^a$, both null vectors being in the equivalence class. 
Again, recall from (\ref{kappa_eq_class}) that the accelerations $\kappa_{(\xi\ell\,)}$
of the null normals $\xi\ell^a$ vary in
the equivalence class through $\kappa_{(\xi\ell\,)}=\xi\kappa_{(\ell\,)}+\lie_\ell\,\xi$.
The point of view here is that the accelerations $\kappa_{(\xi\ell\,)}$ just provides
a nomenclature which can be used as tags to the various surfaces generated by the
corresponding null normals in the equivalence class $[\xi\ell^a]$.
In short, all the null surfaces labelled by say $\kappa_{(\xi_1\ell\,)}$, $\kappa_{(\xi_2\ell\,)}$,
$\kappa_{(\xi_3\ell\,)}$ \emph{etc} are on the equivalent footing from the point of view of boundary
conditions. This exemplifies the claim that the WIH boundary conditions puts the non- extremal
and extremal black hole horizons in the same footing.

An noted several times, the acceleration $\kappa_{(\xi\ell\,)}$  varies over the
class $[\,\xi\ell\,]$: $\kappa_{(\xi\ell\,)}=\xi\kappa_{(\ell\,)}+\lie_\ell\,\xi$ 
and is not a constant in general. In order to obtain the zeroth law,
which requires the acceleration for each normal vector in the class to 
be a constant, we need to restrict the NEHs further. Let us
call the restricted horizon the {\em weak isolated horizon} (WIH), which is 
a NEH equipped with a class $[\,\xi\ell^a\,]$ such that
\beq \label{wih}
\lie_{(\xi\ell\,)}\,\omega^{(\xi\ell\,)}\triangleq 0\;.
\eeq
Let us make a few comments here: First, as in Killing horizons,
we will interpret the acceleration $\kappa_{(\xi\ell\,)}$ as the
surface gravity of $\xi\ell^a$. However, since a global Killing 
field is absent, the value of the surface gravity cannot be
uniquely determined. In isolated horizon formulation it is natural to keep
this freedom. In fact, as we shall see below, this freedom will
enable the extremal horizon with surface gravity $\kappa=0$ and
non- extremal horizons $\kappa\ne 0$ to be on the same footing.  

Second, the boundary condition (\ref{wih}),
unlike the previous ones, is not a {\em single} condition. 
Namely, if it is obeyed by one normal vector $\l^a $  ($\lie_{\ell\,}\,\omega^{(\ell\,)}
\triangleq 0$), then it
is not guaranteed that every other normal vector $\xi\ell^a$ in the class will obey
it. This pathology is eliminated by restricting the choice of $\xi$s
\beq
\xi=c\,e^{-v\kappa_{(\ell\,)}}+\kappa_{(\xi\ell\,)}/\kappa_{(\ell\,)},
\eeq
where,  $c$ is a nonzero function satisfying $\lie_\ell c=0$
and $v$ is the affine parameter such that $\lie_{\l}v=1$.
For the rest of the paper, we choose $c\triangleq$ constant. 
Having restricted ourselves to this specific class it is now easy to
verify that (\ref{wih}) becomes just {\em one} condition:
$\lie_{(\xi\ell\,)}\,\omega^{(\xi\ell\,)}\triangleq 
d(\xi\ell\cdot\omega^{(\xi\ell\,)})\triangleq d\kappa_{(\xi\ell\,)}$ for
every $\xi$ belonging to the restricted class. 
From now on the class of normal vectors 
will always follow this restriction.  This class admits a
$\xi=c\,e^{-v\kappa_{(\ell\,)}}$, for which $\kappa_{(\xi\ell\,)}\triangleq 0$ when
the surface gravity $\kappa_{(\ell\,)}$ of $\ell^a$ is nonzero.
For obvious reasons such an isolated horizon, characterized
by a normal vector of vanishing surface gravity,
will be called an {\em extremal} horizon. Thus, our class of normal vectors
contains both extremal and non-extremal horizons, as opposed to the constant
class of normal vectors $[\,c\ell^a\,]$. In other words, the WIH boundary conditions
cannot differentiate between the surfaces generated by null normals in the equivalence class
$[\xi\ell^a]$ with $\xi=c\,e^{-v\kappa_{(\ell\,)}}+\kappa_{(\xi\ell\,)}/\kappa_{(\ell\,)}$
and hence the extremal and non- extremal horizons become part of the WIH.

As already noted, WIH boundary condition (\ref{wih}) is equivalent to 
the {\em zeroth law}\,: $d\kappa_{(\xi\ell\,)}\triangleq 0$. Therefore, the
surface gravity corresponding to each $\xi\ell^a$ in $[\,\xi\ell^a\,]$
is constant on $\Delta$, provided $\xi$s belong to the restricted class.

\section{The Holst Action}

The Holst action (\cite{Holst}) is a modification of the Palatini action where
a term is added which has a property that it does not contribute
to the equation of motion. Let us recall that the Palatini
action is constructed out of the basic fields, the tetrads $e_{a}^{I}$ 
and a $SO(3,1)$ Lie algebra valued connection one form $A_{IJ}$
(see \cite{Ashtekar_Tate} and \cite{Ashtekar_Lewandowski} for details).
On shell, the connection $A_{IJ}$ equals the spin connection.
The Legendre transformation of Palatini
Lagrangian to the Hamiltonian formulation introduces
second class constraints. Solution of these constraints
needs some gauge fixing which essentially reduces the theory
to that of the standard metric variable theory and one looses the essential  
advantages of the connection formulation. One can however go
to the self dual complex connections where the theory
is much easier but this also creates problems for the quantum theory.
Indeed, the quantum theory based on LQG needs background independent analysis
on the connection space. However, such a theory is still to be
constructed and in the meantime, the quantization programme is successful
on the phase space of the real variables. This phase space is the 
Barbero- Immirzi phase space constructed out of the original Palatini 
phase space by a one parameter canonical transformation labelled by
the Barbero- Immirzi parameter $\gamma$. The Holst action is
precisely the action whose Legendre transformation gives the 
Barbero- Immirzi phase space.

Let us first begin with the Palatini action. We consider the spacetime
$\mathcal{M}$ which is bounded by the the Cauchy surfaces $M_{\pm}$ and
intersecting at $i^0$. For now, we shall only deal with a spacetime without
any inner boundary $\Delta$. The action is given by:
\begin{figure}[h] \label{f1}
  \begin{center}
  \includegraphics[height=4.0cm]{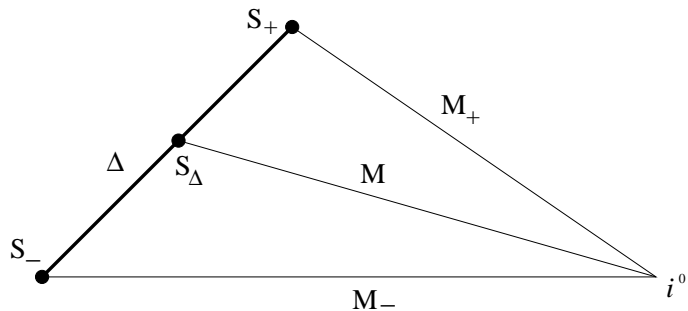}
  \caption{$M_\pm$ are two partial Cauchy surfaces enclosing a  region of
space-time and intersecting $\Delta$ in the $2$-spheres $S_\pm$ respectively
and extend to spatial infinity $i^o$. Another Cauchy slice M is drawn which
intersects $\Delta$ in $S_\Delta$}
  \end{center}
\end{figure}

\begin{eqnarray}\label{Palatini_action} S_{P}&=&\dfrac{1}{32\pi
G}\int_{\mathcal{M}}\epsilon_{IJKL}e^I \wedge e^J\wedge F^{KL}\nn
&=&\dfrac{1}{16\pi G}\int_{\mathcal{M}}\Sigma_{IJ}\wedge F^{IJ}
\end{eqnarray}
where, $\Sigma_{IJ}=\frac{1}{2}\epsilon_{IJKL}e^K \wedge e^L$ is
a two- form 
defined for future convenience.
The variation of the integral with respect to $A_{IJ}$ will lead to :
\begin{equation} \delta S_{P}=\dfrac{1}{16 \pi G}\int_{M} d(\Sigma_{IJ}\wedge
\delta A^{IJ}) -\dfrac{1}{16 \pi G}\int_{M} D\Sigma_{IJ}\wedge \delta A^{IJ}
\end{equation}
where we have used the fact that:
\begin{equation}
D\Sigma_{IJ}:=d\Sigma_{IJ}+A_{I}{}^{K}\wedge\Sigma_{KJ}+A_{J}{}^{K}\wedge\Sigma_{IK}
\end{equation}
Thus, the equation of motion obtained from variation of the connection is :
\begin{equation}\label{dsigma1} D\Sigma_{IJ}=0
\end{equation}
It is then easy to check from (\ref{dsigma1}) that the connection $A_{IJ}$ is
the spin connection. Then, the variation of the action (\ref{Palatini_action}) with
respect to the tetrads $e_{I}^{a}$ gives the Einstein equation. In the present case,
observe that the boundary term goes to zero
at the outer boundary where the spacetime is flat and the variational principle
is well defined. In cases where there is an inner boundary, we need to take care
that the boundary term again vanishes.

Now, let us consider the Holst action \cite{Holst}:
\begin{equation} S_{H}= S_{P}- \dfrac{1}{16 \pi G \gamma}\int_{M}e^I \wedge
e^J\wedge F_{IJ}
\end{equation}
where, $\gamma$ is a real constant called the Barbero -Immirzi parameter. Note
that written in this form, the parameter cannot be equal to zero. For
$\gamma=-i~ (+i)$, we get the self dual (anti self-dual) action. Then, the
action $S_{H}$ can be written in terms of
${}^{\pm}F_{IJ}(A_{IJ})=F_{IJ}({}^{\pm}A_{IJ})$. In these cases, it can be
argued that the connection is actually a self-dual (anti-self dual) part of
the spin connection. For generic but other values of $\gamma$, it is also true
that the connection $A_{IJ}$ is actually a spin connection, completely
determined by the tetrad.  The variation of the 
action with respect to the tetrad again reproduces the
Einstein equation, the $\gamma$- dependent term, on shell,
being zero by the algebraic Bianchi identity.

The extra $\gamma$- dependent
term that appears in the action
is important for quantum mechanical reasons. It is in some sense similar to the
\emph{theta} ($\theta$) term in QCD. It is well known that in Yang- Mills
theory, $\theta-$ term introduces inequivalent  $\theta-$\emph{sectors} for
the corresponding quantum theory. In the similar way, the extra term here
introduces the inequivalent $\gamma-$\emph{sectors} for quantum general
relativity \cite{Rovelli_Thiemann}. However, while the $\theta
(\bfF_{YM}\sbfF_{YM})$ is a purely topological term\footnote{In some String
theories, like the $E_{8}\otimes E_{8}$ heterotic string theory, the $\theta-$
term arises  from the Yang- Mills Chern-Simons term which makes the theory
gauge anomaly free. In that case, derivative of $\theta$ is a field dual to
the Kalb- Ramond three form.  We consider the case where $\theta$ in a
constant just like in ordinary QCD.}, being equal to a total divergence of a
$3-$form, the  $\gamma-$ dependent term here vanishes because of the first
Bianchi identity. One can show that the phase space corresponding to 
these two theories are equal. To see this one can construct the symplectic structure
for the Holst action and argue that if the phase space vectors satisfy the
linearized version of the spin connection equation, the $\gamma$- dependent
term disappears (see appendix \ref{sec:Holst_Vs_Palatini} for the details).
In other words, the Holst modification of the Palatini action
implies a canonical transformation on the phase space.


\subsection{The Action  and the  Variational Principle}\label{subsec:variation}
In this subsection, we will use the Holst action for a spacetime
with inner boundary. In the present case, the inner boundary is a null
surface which satisfies the WIH boundary conditions. To be more precise,
the spacetime under consideration is a region
bounded by the Cauchy surfaces $M_{1}$ and $M_{2}$ extending
to spatial infinity and the null surface $\Delta$ (see fig. 1). The variation
of the fields will be between all those configurations which
satisfy the boundary conditions at infinity and at $\Delta$. In particular,
we consider those variations of ($e^I{}_{a}, A_{IJ}$) which satisfy the 
standard fall off conditions at infinity and on  $\Delta$, satisfy the
following conditions:
\begin{enumerate}
\item each spacetime admits a null normal belonging to the equivalence class $[\xi\l^a]$.
\item each pair ($\Delta, [\xi\l^a ]$) is a WIH.
\end{enumerate}

We can always introduce a fixed set of internal null vectors $(\l^{I}, n^{I}, m^{I}, \bar m^{I})$
on $\Delta$ such that $\partial_{a}(\l^{I}, n^{I}, m^{I}, \bar m^{I})=0$ (this partially fixes
the Lorentz frame). Given these internal 
null vectors and the tetrad $e^I{}_{a}$, we can construct the null vectors $(\l^{a}, n^{a}, m^{a}, \bar m^{a})$
through $\l_a =e^I{}_{a}\l_{I} $.

\subsubsection*{Tetards and Connection on $\Delta$}\label{subsec:connection} 
To proceed further, we need the the expressions for the
tetrad and the connection in terms of the null vectors.
The expansion of tetrads in terms of the
null vectors can be easily calculated from the
expression of tetrads relating the spacetime metric $g_{ab}$ and 
internal flat metric $\eta_{IJ}$. The expression is given by:
\begin{equation}\label{tetrad_exp}
e^{I}{}_{\ub{a}}\= -n_{a}\l^{I} + m_{a}\bar m^{I} + \bar m_{a} m^{I}
\end{equation}
The above equation (\ref{tetrad_exp}) can be used to get an expression for the 
product of tetrads:
\begin{eqnarray}\label{tetrad_prod_exp}
e^{I}{}_{\ub{a}}\wedge e^{J}{}_{\ub{b}} &\=& -2~n_{a}\wedge m_{b}~\l^{[I}\bar m^{J]} -2~n_{a}\wedge \bar m_{b}~\l^{[I} m^{J]} +2 m_{a}\wedge\bar m_{b}~\bar m^{[I}m^{J]}\\ \nn 
&=&-2~n_{a}\wedge m_{b}~\l^{[I}\bar m^{J]} -2~n_{a}\wedge \bar m_{b}~\l^{[I} m^{J]} +2i~m^{[I}\bar m^{J]}~{}^2\mbf{\epsilon}_{ab}
\end{eqnarray}
Using this expression for the tetrad products (\ref{tetrad_prod_exp}), and the expansion
for the internal epsilon tensor $\epsilon_{IJKL}=4!\l_{[I}n_{J}m_{K}\bar m_{L]}$, we get
\begin{equation}
\Sigma_{\ub{ab}}{}^{IJ}\= 2\l^{[I}n^{J]}~ {}^2\mbf{\epsilon}_{ab} +2n_{a}\wedge(im_{b} \l^{[I}\bar m^{J]} - i\bar m_{b}\l^{[I} m^{J]})
\end{equation}

We are now in a position to calculate the connection
$A_{IJ}$.  We will be using the Newman- Penrose formalism. The details
of the Newman- Penrose coefficients are given in the appendix \ref{appb}. Using those
expansions, one gets the following expression for the covariant derivatives
of the null normals pulled back and restricted to $\Delta$.

\begin{equation} \nabla_{\ub{a}}\l^{b}\=\omega^{(\ell\,)}_{a}\l^{b}
\end{equation}
\begin{equation}\label{exp_normal_n}
\nabla_{\ub{a}}n^{b}\=-\omega^{(\ell\,)}_{a} n^{b}+\bar{U}^{(l,m)}_{a}m^{b}+U^{(l,m)}_{a} \bm^{b}
\end{equation}
\begin{equation} \nabla_{\ub{a}}m^{b}\={U^{(l,m)}}_{a}\l^{b}+V^{(m\,)}_{a}m^{b}
\end{equation}
\begin{equation} \nabla_{\ub{a}}\bm^{b}\=\bu^{(l,m)}_{a}\l^{b}-V^{(m\,)}_{a}\bm^{b}
\end{equation}
where, the superscripts for each of the one forms keep track of their
dependencies on the 
rescaling of the corresponding null normals. The expressions of the one forms $\omega^{(\ell\,)}, U^{(l,m)}, \bu^{(l,m)}$ and $V^{(m)}$ can
be written in terms of the null normals and are as follows:
\begin{eqnarray} \omega^{(\ell\,)}_{a}&\=& -\left( \epsilon + \bepsilon\right)n_{a}+\left(
\balpha+\beta \right)\bm_{a}+\left( \alpha+\bbeta \right)m_{a}\nn 
U^{(l,m)}_{a} &\=&-\bpi n_{a}+\bmu m_{a} +\blambda \bm_a\nn
 V^{(m\,)}_{a}&\=&-\left( \epsilon-\bepsilon\right)n_{a}+\left( \beta -\balpha \right)\bm_{a}+\left(\alpha-\bbeta \right)m_{a}
\end{eqnarray}
The part of the connection $V^{(m\,)}$ is purely imaginary. Let us at this stage
point out the result of the rescaling of the null normal $\l^a $ on the
various quantities of interest.
Firstly, for $\l^a\longrightarrow  \xi\l^a $ , we have:
\begin{equation}
\omega^{(\ell\,)}_{a}\rightarrow \omega^{(\xi\ell\,)}_{a}=\omega^{(\ell\,)}_{a}+\nabla_{a}ln \xi
\end{equation} 

Since the normalization of $\l^a$ and $n^a$ are connected, we must have $n^a\longrightarrow  \frac{n^a}{\xi} $ when $\l^a\longrightarrow  \xi\l^a $. Then the effect of the rescaling can be seen to be: 
\begin{equation}
\nabla_{\ub{a}}\left( \frac{n^{b}}{\xi}\right) \=-\omega^{(\xi\ell\,)}_{a}\left( \frac{n^{b}}{\xi}\right) + \bar{U}^{(\xi\l,m)}_{a}
m^{b}+ U^{(\xi\l,m)}_{a}\bm^{b}
\end{equation}
Thus, under this transformation, we have that the one- form $ \omega^{(\xi\ell\,)}$ transforms
in the usual way: $ \omega^{(\ell\,)}\rightarrow \omega^{(\xi\ell\,)}_{a}=\omega^{(\ell\,)}_{a}+\nabla_{a}ln \xi $
and the other one forms $U^{(l,m)}$ and $\bar U^{(l,m)}$  transform as
\begin{eqnarray} 
\bar{U}^{(\l,m)}_{a}\longrightarrow \bar{U}^{(\xi\l,m)}_{a}=\frac{\bar{U}^{(\l,m)}_{a}}{\xi}\\ \nonumber
U^{(\l,m)}_{a}\longrightarrow U^{(\xi\l,m)}_{a}=\frac{U^{(\l,m)}_{a}}{\xi}
\end{eqnarray}

This rescaling is in the sector of $\l, n$. There can be another set of
rescaling quite independent of the rescaling of $\l, n$. This
concerns the transformation in the other set of null vectors $m,\bar m$ of the null tetrad.
This transformation function will also be
independent of the function $\xi$.  Now, for $m\rightarrow f m$ and for $\bm\rightarrow \frac{\bm}{f}$, 
where $f$ is any function on $\Delta$, we have the following transformations:
\begin{equation}
\nabla_{\ub{a}}\left(f m^{b}\right)\= U^{(\l,fm)} _{a}\l^{b}+V^{(fm\,)}_{a}\left(fm^{b}\right)
\end{equation}
%
%
\begin{equation}
\nabla_{\ub{a}}\left(\frac{\bm}{f}\right)\=\bar{U}^{(\l,fm)}_{a}\l^{b}-V^{(fm\,)}_{a}\left(\frac{{\bm}^{b}}{f}\right)
\end{equation}
The transformation rules are as follows for the one forms $U^{(\l,m)}_{a}$, $\bar{U}^{(\l,m)}_{a}$ and $V^{(m\,)}_{a}$ are
as follows:
\begin{eqnarray}\label{trans_rule_uuv} 
\bar{U}^{(\l,m)}_{a}&\longrightarrow &\bar{U}^{(\l,fm)}_{a}=\frac{\bar{U}^{(\l,m)}_{a}}{f}\\ \nonumber
U^{(\l,m)}_{a}&\longrightarrow & U^{(\l,fm)}_{a}=f U_{a}\\ \nonumber
V^{(m\,)}_{a}&\longrightarrow & V^{(fm\,)}_{a}=V^{(m\,)}_{a}+ \nabla_{a} ~lnf
\end{eqnarray}

The part of the connection $\omega^{(\l)}$ and $V^{(m\,)}$ transform
as $U(1)$ field whereas the the other parts of connections 
only rescale. 
We have constrained only one part of the connection
while defining the Weak Isolated Horizon in the sense that
only the one form $\omega$ is constrained. The other part of the connection is left 
as it is. If we want to constrain more of the parts of the
connection then we get the definition of the Isolated horizon.

We can use these information to find the connection. 
To do this, we first note that the internal null vectors
are fixed such that $\partial_{a}(\l^{I}, n^{I}, m^{I}, \bar m^{I})=0$.
Then, we get 
\begin{equation}\label{eq-conn-cal}
\nabla_{\ub{a}}\l_{I}\= A_{\ub{a}I}{}^{J}\l_{J}
\end{equation}
We can choose a tetrad $e_{a}^{I}$ which maps the vector $\l_{I}$ to $\l_{a}$.
This tetrad is annihilated by the covariant derivative, $\nabla_{a}e_{b}^{I}=0$. 
Then the equation (\ref{eq-conn-cal}) gives: $A_{\ub{a}}{}^{I}{}_{J}\l^{J}\=\omega_{a}^{(\l)}\l^{I}$.
Written is a more compact form, this reduces to:
\begin{equation}
 A_{\ub{a}IJ}\=-2\omega_{a}^{(\l)}~\l_{[I}n_{J]} + Q_{IJ}
\end{equation}
where, the one form $ Q_{IJ}$ is such that $ Q_{IJ}\l^{J}\=0$.

We can proceed just as before for the other null vectors $n_{I},~ m_{I}$ and $\bar m_{I}$. For the null vector $n_{I}$,
that gives us:
\begin{equation}
 A_{\ub{a}IJ}\=-2\omega_{a}^{(\l)}~\l_{[I}n_{J]} -2\bar U^{(\l,m)}_{a}~m_{[I}\l_{J]}-2U^{(l,m)}_{a}~\bar m_{[I}l_{J]} +  R_{IJ}
\end{equation}
where, the one form $ R_{IJ}$ is such that $ R_{IJ}n^{J}\=0$. For the null vector $m_{I}$, we get:
\begin{equation}
 A_{\ub{a}IJ}\= -2U^{(\l,m)}_{a}~\bar m_{[I}l_{J]}+ 2V_{a}^{(m)}~m_{[I}\bar m_{J]} +  S_{IJ}
\end{equation}
where, the one form $ S_{IJ}$ is such that $ S_{IJ}m^{J}\=0$. A similar construction for the
null vector $\bar m_{I}$ implies
\begin{equation}
 A_{\ub{a}IJ}\= -2\bar U^{(\l,m)}_{a}~\bar m_{[I}l_{J]}+2V_{a}^{(m)}~m_{[I}\bar m_{J]} + \bar S_{IJ}
\end{equation}
such that $ \bar S_{IJ}\bar m^{J}\=0$

The connections that we have obtained above complement each other.
Combining all these expressions, we get the complete expression for the
connection $A_{\ub{a}IJ}$. These gives the connection to be:
\begin{equation} A_{IJ}=-2~\omega^{(\l)}~ \l_{[I}n_{J]}+ 2~U^{(l,m)}~\l_{[I}\bm_{J]}+ 2~{\bar U}^{(l,m)}~\l_{[I}
m_{J]}+ 2~V^{(m)}~m_{[I}\bm_{J]}
\end{equation}
We define the following connection for ease of computation \footnote{This choice of the connection $A^{(H)}_{IJ}$ with a factor $\frac{1}{2}$ in front is made to make our results
conform to the standard results \cite{abck}. However, this factor can be arbitrarily
chosen and the quantum implementation of the boundary conditions including state counting goes through unchanged entirely.}
\begin{equation} 
A^{(H)}_{IJ}:=\frac{1}{2}\left( A_{IJ}-\frac{\gamma}{2}
\epsilon_{IJ}{}^{KL}A_{KL}\right)
\end{equation}

This leads to the following form of the connection:
\begin{eqnarray} A^{(H)}_{aIJ}&\=& ~\l_{[I}n_{J]}~\left( -\omega^{(\l)}_{a}+i\gamma V^{(m)}_{a}
\right) + ~ m_{[I}\bm_{J]}~\left(V^{(m)}_{a}-i\gamma \omega^{(\l)}_{a} \right)\nn
&+& ~\l_{[I}\bm_{J]}~\left( U^{(l,m)}_{a}+i\gamma U^{(l,m)}_{a}
\right)+ ~\l_{[I}m_{J]}~\left(\bu^{(l,m)}_{a}-i\gamma \bu^{(l,m)}_{a} \right)
\end{eqnarray}

\subsubsection*{Variation of the Action}
The next step is to check the variational principle. The lagrangian that we 
are interested in is of the form:
\begin{equation}
-16\pi G\gamma~L= \gamma \Sigma_{IJ}\wedge F^{IJ}~-~e_{I}\wedge e_{J}\wedge F^{IJ}~ -~\gamma~d(\Sigma_{IJ}\wedge A^{IJ}) +~d(e_{I}\wedge e_{J}\wedge A^{IJ}).
\end{equation}
where we have added the two boundary term just for convenience. These terms will not contribute
to the equation of motion.
The variation of the action on-shell will give two terms on the 
boundary $\Delta$. They are:
\begin{equation}
\delta S (e,A) = \frac{-1}{8\pi G \gamma}\int_{\Delta}(iV^{(m)} + \gamma~\omega^{(\l)})\wedge\delta{}^2\mbf{\epsilon}
\end{equation}

We can argue that the term is zero and hence the action principle is well defined.
The argument goes as follows. First of all, the field configurations
over which the variations are taken are such that they satisfy the standard
boundary conditions at infinity and the WIH boundary conditions at $\Delta$.
The weak isolation condition implies that $\lie_{\l}\omega^{(\l)}\=0$
though there is no such condition on $V^{(m)}$. However 
interestingly, $d\omega^{(\l)}$ and $dV^{(m)}$ are proportional
to ${}^2\mbf{\epsilon}$ and hence inner product with $\l^a$ of these 
quantities are zero. This implies that for variations among
field configurations with null normals in the equivalence class,
we have $\lie_{\xi\l}\omega^{(\l)}\= d(\xi\kappa_{(\l)})$ and $\lie_{\xi\l}V^{(m)}\= d(\xi(\epsilon-\bar\epsilon))$. 
This implies that on the application of  $\lie_{\xi\l}$, the integral goes to the initial
and the final cross section of $\Delta$. However, the variation of the fields for example $\delta{}^2\mbf{\epsilon}$
is zero at the initial and final hypersurface by the standard rules of variational principle. Thus the integral is lie dragged by any null normal in the equivalence class.
In other words, the integral in zero at the initial and the final hypersurface and is lie dragged
on $\Delta$ . Thus, the entire integral is zero and the action principle is well defined.

\subsection{The Symplectic Structure}
\label{sec:Symplectic}
The phase space for the system can be constructed. 
We recall that the variation of the Lagrangian produces the three form $\Theta(\delta)$, such that
$\delta L=: d\Theta (\delta)$. In the present case, we have:
\begin{equation}
16\pi G\gamma~\Theta(\delta)=\gamma~\delta\Sigma_{IJ}\wedge A^{IJ}-\delta(e_{I}\wedge e_{J})\wedge A^{IJ}=-2~\delta(e^{I}\wedge e^{J})\wedge A^{(H)}_{IJ}
\end{equation}
The construction of the symplectic current from here is standard.
The current is $J(\delta_{1}, \delta_{2}):=\delta_{1}\Theta(\delta_{2})- \delta_{2}\Theta(\delta_{1})$.
The current is closed on- shell \emph{i.e.} $dJ=0$.
 The resulting Symplectic Current is :
\begin{equation} J\left( \delta_1, \delta_2
\right):=\dfrac{1}{8\pi G\gamma}\left\lbrace \delta_{[1}\left( e_{1}\wedge
e_{2}\right) \right\rbrace \wedge\left\lbrace \delta_{2]}\left(
A_{IJ}-\frac{\gamma}{2} \epsilon_{IJ}{}^{KL}A_{KL}\right)\right\rbrace
\end{equation}
Integrating the symplectic current over $\mathcal{M}$, we get the contribution of the symplectic current
from the boundaries of the spacetime region under consideration:
\begin{equation}
\int_{M_{+}\cup M_{-}\cup\Delta\cup i^{0}}J(\delta_{1}, \delta_{2})=0
\end{equation}

The boundary conditions at infinity ensure that the integral of the symplectic current
at $i^0$ vanishes. However, to construct the symplectic structure, we must be
careful that no data flows out of the phase space because of our choice of foliation.
To ensure this, we must check that the symplectic structure is independent of the
choice of foliation. For that, we introduce potentials:
\begin{enumerate}
\item $\lie_{(\xi\l)}\psi_{(\xi\l)}\=\xi\l^a \omega^{(\xi\l)}_{a}\=\kappa_{(\xi\l)}$
\item $\lie_{(\xi\l)}\mu_{(m)}\=i\xi\l^a V^{(m)}_{a}\= i\xi(\epsilon-\bar\epsilon)$
\end{enumerate}
along with the boundary condition that they are zero at one of the cross- sections
of $\Delta$ so as to fix the additive ambiguities. We choose $\psi_{(\xi\l)}=0$ and $\mu_{(m)}=0$ at $S_{-}$.

The basic idea now is to write the symplectic current $J(\delta_{1}, \delta_{2})$ on $\Delta$
in terms of these potentials and see that $J(\delta_{1}, \delta_{2})\= dj(\delta_{1}, \delta_{2})$.
With this simplification, the integrals of $J(\delta_{1}, \delta_{2})$ on $\Delta$
will be taken to the boundaries $S_{\pm}$ of $\Delta$. 
To see this, first note that the expression of symplectic current $J(\delta_{1}, \delta_{2})$ on $\Delta$
is given by:
\begin{equation}
J(\delta_{1}, \delta_{2})|_{{}_{\Delta}}\= \frac{-1}{8\pi G \gamma}[\delta_{1}{}^2\mbf{\epsilon} \wedge \delta_{2}(iV+\gamma\omega^{(\l)})- \delta_{2}{}^2\mbf{\epsilon} \wedge \delta_{1}(iV+\gamma\omega^{(\l)})]
\end{equation}
The potential for $\kappa_{(\l)}$ is $\psi_{(\l)}$. It can be seen from the definition that
$\psi_{(\l)}$ is a function of $v$ only. The potential for $i(\epsilon-\bar\epsilon)$
is $\mu_{(m)}$ and it can be seen again that the wedge product of the variation of $\epsilon$
and variation of $d\mu_{(\l)}+\lie_{\l}\mu_{(m)}n$ vanish. These two results imply that
\begin{equation}\label{exp_j_delta_d}
J(\delta_{1}, \delta_{2}))|_{{}_{\Delta}}\=d\left[\frac{-1}{8\pi G\gamma} \left(\delta_{1}{}^2\mbf{\epsilon}~\delta_{2}(\mu_{(m)} +\gamma\psi_{(\l)})- (1\leftrightarrow 2)\right)\right]
\end{equation}

We take a particular orientation of the spacetime foliation into account. 
That gives us the result that the symplectic current is independent of the foliation.
\begin{equation}
(\int_{M_{+}}- \int_{M_{-}})J(\delta_{1}, \delta_{2})\=\frac{1}{8\pi G\gamma}(\int_{S_{-}}-\int_{S_{+}})\{\delta_{1}{}^2\mbf{\epsilon}~\delta_{2}(\mu_{(m)} +\gamma\psi_{(\l)})- (1\leftrightarrow 2)\}
\end{equation}
The construction of symplectic current is independent of our choice foliation
and hence all the phase space information can be obtained from this symplectic current
by staying on any arbitrary foliation. 
We choose a particular Cauchy surface $M$ which intersects $\Delta$
in the sphere $S_{\Delta}$.
 The symplectic structure is then given by :
\begin{eqnarray}\label{Palatini_1}
\Omega(\delta_{1}, \delta_{2}
)&:=&\frac{1}{8\pi G\gamma}\int_{M}\left[ \delta_{1}(e^{I}\wedge
e^{J})~\wedge\delta_{2}A^{(H)}_{IJ} -\delta_{2}(e^{I}\wedge
e^{J})~\wedge\delta_{1}A^{(H)}_{IJ} \right] \nn
&+&\frac{1}{8\pi G\gamma}\int_{S_{\Delta}}\left[
\delta_{1}{}^2\mbf{\epsilon}~\delta_{2}(\mu_{(m)}+ \gamma\psi_{(\l)}) -
\delta_{2}{}^2\mbf{\epsilon}~\delta_{1}(\mu_{(m)} + \gamma\psi_{(\l)})\right]
\end{eqnarray}

\section{The First Law }
\label{sec:phase_space} 
The first law requires defining an energy. Since the WIH is
a local definition of a horizon, the first law should involve only locally defined quantities.
To be more precise, the first law is expected to relate variations
of local quantities that are defined only at the horizon without any reference to the rest of the spacetime.
We already have the surface gravity $\kappa_{(\xi\l)}$ defined only
locally at the horizon and the other quantity that we require now is a locally defined
energy (for horizons carrying other charges, such as angular momentum, electric potential etc., we must also provide local definitions for them). To proceed, it should be noted that in spacetime, energy is associated
with a timelike Killing vector field. Given any vector field $W$
in spacetime, it naturally induces a vector field $\delta_{W}$ in the phase space.
The phase space vector field $\delta_{W}$ is the generator of time translation in
the phase space. If time translation is a canonical transformation in the phase space then
$\delta_{W}$ defines a Hamiltonian function $H_{W}$ for us.

So to find out the Hamiltonian function associated with energy, we must look for phase space transformations that keep the symplectic structure invariant, in other words the canonical
transformations. The vector fields tangent to these canonical flows are the Hamiltonian vector fields.
To check wheather a vector field $\delta_{t}$ in the phase space is Hamiltonian, one constructs a one-form $X_t$ where $X_{t}(\delta):=\Omega(\delta,\delta_{t})$, where $\delta_{t}$ is the lie flow $\lie_{t}$ generated by the spacetime vector field $t^a$ when tensor fields are varied. The necessary and sufficient
condition for the vector field $\delta_{t}$ to be a \emph{globally Hamiltonian} vector
field is that the one-form $X_{t}$ is to be \emph{exact}, $X_{t}=\mathbf{d}H_{t}$
where $\mathbf{d}$ is the exterior derivative in phase space
and $H_t$ is the corresponding Hamiltonian function. In other words, the vector field $\delta_{t}$
is globally Hamiltonian if and only if $X_{t}(\delta)=\delta H_{t}$ for any vector field $\delta$
in the phase space. Because of the presence of the boundary, the WIH, The vector fields $t^a$ are also
restricted by the condition that it should be tangential on $\Delta$. Now being a null surface, the WIH 
has only three tangential directions, one null and the two other spacelike. The closest analog of `time' 
translation on WIH is therefore translation along the null direction. It is generated by the vector field
$[\xi\l^a]$ (one can take any but fixed $\xi$ belonging to the equivalence class). For global solutions 
this null normal vector field becomes timelike outside the horizon and is expected to match with the asymptotic
time-translation for asymptotically flat spacetimes. So in this sense, the local energy on the horizon is the local snapshot of ADM or Bondi energy defined at spacelike/null asymptotic infinity.

The relevant question is: Is the flow generated by the phase space vector field $\delta_{\xi\l}$
Hamiltonian? To find that, we calculate the symplectic structure for any arbitrary live vector field
$\delta_{\xi\l}$. It is useful to recall that the action of the phase space vector field  $\delta_{\xi\l}$
on tensor fields is the lie flow $\lie_{\xi\l}$ generated by the vector field $\xi\l^a$. For the above symplectic structure, $X_{(\xi\l)}(\delta)$ gets contribution from both the bulk and the surface symplectic structure. The bulk term, thanks to
the equation of motion satisfied by the fields and their variations, contributes
only through the boundaries of the Cauchy surface $M$, which are the $2-$ spheres $S_{\Delta}$
and $S_{\infty}$ respectively: 

\begin{equation}\label{bulk_X}
X_{\xi\l}(\delta)|_{M}=\frac{-1}{8\pi G}\xi\kappa_{(\l)}\delta \mathcal{A}_{\Delta} -\frac{i}{8\pi G\gamma}\int_{S_{\Delta}}\xi(\epsilon -\bar\epsilon)~\delta {}^2\mbf\epsilon +\delta E_{(\xi\l)}
\end{equation}
where, $\mathcal{A}_{\Delta}=\int_{S_{\Delta}}\epsilon$ is the area of 
$S_{\Delta}$ and $E_{(\xi\l)}$ is the ADM energy arising 
out of the integral at $S_{\infty}$, assuming that the asymptotic time 
translation matches with the vector field $\xi\l^a$ at infinity.

The  $X_{(\xi\l)}(\delta)$ also gets contribution from the 
surface symplectic structure.
For $t=\xi\l^a$, we must be careful with the evaluations of
the action of $\delta_{(\xi\l)}$ on the potentials $\mu_{(m)}$
and $\psi_{(\l)}$. The action of $\delta_{(\xi\l)}$ cannot
be interpreted as $\lie_{(\xi\l)}$ when acting on potentials.
To determine the action, we proceed as follows.
For the case of $\psi_{(\l)}$, it is clear that since
variation of $\psi_{(\xi\l)}$ is completely determined by
$\kappa_{(\xi\l)}$, $\delta_{(\xi\l)}\psi_{(\xi\l)}=0$.
However, $\psi_{(\xi\l)}=\psi_{(\l)}+ln~\xi $ implies that
$\delta_{(\xi\l)}\psi_{(\l)}=-\lie_{\l}\xi$.
For the other potential, observe that $\delta_{(\xi\l)}\mu_{(m)}-i(\epsilon-\bar\epsilon)$
satisfies the differential equation $\lie_{\xi\l}(\delta_{(\xi\l)}\mu_{(m)}-i(\epsilon-\bar\epsilon))=0$
with the boundary condition that $\mu_{(m)}=0$ at the point $v=0$. This implies that
because $(\epsilon-\bar\epsilon)=0$ at $v=0$, the action is $\delta_{(\xi\l)}\mu_{(m)}=i(\epsilon-\bar\epsilon)$. The considerations above leads to:
\begin{eqnarray}\label{bndy_X}
X_{\xi\l}(\delta)|_{S_{\Delta}}&=&
\frac{1}{8\pi G\gamma}\int_{S_{\Delta}}[\delta{}^2\mbf\epsilon~\delta_{\xi\l}(\mu_{(m)}+ \gamma\psi_{(\l)})-\lie_{\xi\l}{}^2\mbf\epsilon~\delta(\mu_{(m)}+ \gamma\psi_{(\l)})] \\ \nonumber
&=&-\frac{1}{8\pi G}\lie_{\l}\xi~\delta\mathcal{A}_{\Delta}+\frac{i}{8\pi G\gamma}\int_{S_{\Delta}}\xi(\epsilon -\bar\epsilon)~\delta{}^2\mbf\epsilon \\ \nonumber
\end{eqnarray}

Combining the two equations (\ref{bulk_X}) and (\ref{bndy_X}), we get:
\begin{equation}
X_{\xi\l}(\delta)\=-\frac{1}{8\pi G}\kappa_{(\xi\l)}\delta \mathcal{A}_{\Delta}+ \delta E_{(\xi\l)}
\end{equation}
This is a fundamental result of the generalization to the most general class of
null normals $[\xi\l^a]$. In the constant class of null normals, there
is no contribution from the surface symplectic structure. In the
 generalized class of null normals, the precise contribution
from the bulk (\ref{bulk_X}) and the boundary (\ref{bndy_X}) leads to
the physically meaningful variation.

The condition that $\delta_{\xi\l}$ is a Hamiltonian
implies that the surface gravity $\kappa_{(\xi\l)}$ is a function
of area $\mathcal{A}_{\Delta}$ only. 
To see that note that to check that $\delta_{\xi\l}$ is a Hamiltonian,
we must check that $\mathbf{d}X_{\xi\l}=0$. In other words, this 
implies $\mathbf{d}X_{\xi\l}(\delta_{1}, \delta_{2})=0$. A simple
calculation gives:
\begin{eqnarray}
\mathbf{d}X_{\xi\l}(\delta_{1}, \delta_{2})&=&\delta_{1}(\kappa_{(\xi\l)}~\delta_{2}\mathcal{A}_{\Delta})-\delta_{2}(\kappa_{(\xi\l)}~\delta_{1}\mathcal{A}_{\Delta})\\ \nn
&=&\delta_{1}\kappa_{(\xi\l)}~\delta_{2}\mathcal{A}_{\Delta}-\delta_{2}\kappa_{(\xi\l)}~\delta_{1}\mathcal{A}_{\Delta}
\end{eqnarray}
This can be written in a more suggestive form as $\mathbf{d}\kappa_{(\xi\l)}\wedge\mathbf{d}\mathcal{A}_{\Delta}(\delta_{1}, \delta_{2})=0$. This implies that since $\delta_{1}$ and $\delta_{2}$ are arbitrary, the wedge product 
is zero by itself \emph{i.e.} we get:
\begin{equation}
\mathbf{d}\kappa_{(\xi\l)}\wedge\mathbf{d}\mathcal{A}_{\Delta}=0
\end{equation} 
which implies that the surface gravity $\kappa_{(\xi\l)}$ is a function
of area $\mathcal{A}_{\Delta}$ only. The exact functional form however remains
undetermined. 
This also implies that there exists
a locally defined function $ E_{\Delta}$. 
Defining the total Hamiltonian $X_{\xi\l}(\delta)=\delta H_{\xi\l}=: E_{(\xi\l)}- E_{\Delta}$, we get the first law of weak isolated
horizons:
\begin{equation}
\delta E_{\Delta}\= \frac{1}{8\pi G}\kappa_{(\xi\l)}~\delta \mathcal{A}_{\Delta}
\end{equation} 
This is consistent since the previous condition imples the existence
of a locally defined energy such that $\mathbf{d}\kappa_{(\xi\l)}\wedge\mathbf{d}\mathcal{A}_{\Delta}(\delta_{1}, \delta_{2})=0$.
With the present first law, we see that this relationship holds.  

\section{Chern-Simons Theory from Symplectic Structure}
Once the four laws of black hole mechanics are established one wonders
whether their resemblances with the four laws of thermodynamics are
pure coincidences? In a brilliant paper, Bekenstein argued that this
is not a mere resemblance but black holes indeed may have entropy
proportional to the classical area of the horizon. Shortly after, by
analyzing quantum fields in a collapsing spacetime Hawking showed that
a black hole has a temperature of (in units of $\hbar=k_B=1$)
$\kappa/2\pi$, where $\kappa$ is the surface gravity at the
horizon. It readily implies that under semiclassical approximation the
entropy of a black hole is equal to the one-quarter of its horizon
area. This interpretation of entropy as area obviously begs for its
derivation from first principles, namely the Boltzmann definition of
entropy arising out the microstates of the system. However, in the
present case the system being a black hole spacetime, the question is
what are its microstates? In other words, what are the microscopic
constituents of a black hole spacetime to which such microstates are
to be assigned? This is indeed a deep question and its answer lies
beyond the general theory of relativity. Since the laws of the
microscopic world are quantum mechanical, one naturally asserts that
such microscopic constituents of spacetime obey the laws of quantum
mechanics, rather than the classical laws of general relativity. This
therefore, warrants for a quantum theory of spacetime or a quantum
theory of gravity. Such a theory, to our complete satisfaction, is
simply not available at present.

One general approach towards a statistical interpretation of black
hole entropy is the {\em loop approach} which flourished under the
umbrella of loop quantum gravity. In this approach, the statistical
analysis of the microscopic constituents for a generic black hole was
first presented in a series of papers \cite{abck, abk, ack}. The
general idea was to not to find out the microscopic constituents of an
entire black hole spacetime, but rather their imprints on the
classical horizon of the black hole, which when treated as a boundary
of the spacetime outside of a black hole, gets excited by some
effective degrees of freedom that arise due to a delicate but
well-defined interaction between the boundary and the bulk of the
spacetime. One assumes that these effective degrees of freedom capture
the bare minimal features of a black hole spacetime, thereby it is
only natural that such effective states are localized only at the
horizon rather than spread out all over the spacetime. The isolated
horizons become relevant in this context because such surfaces are
tailored to capture the essential features of a black hole
spacetime. One then quantizes this effective theory induced at an
isolated horizon and and count the appropriate quantum states. This
turn out to be consistent with the semiclassical estimates made by
Bekenstein and Hawking. Furthermore, the effective theory on the
horizon can only be a theory of the topological kind, namely it must
be insensitive to the metric on the horizon. This is because the
horizon is a null surface and therefore cannot support a physical
particle. The above papers show, through a detailed canonical phase
space analysis, that the effective theory on the horizon is
Chern-Simons type, more precisely a $U(1)$ Chern-Simons theory.

The main objective of the section is to find out the effective field
theory on a spherically symmetric WIH, starting from the Holst action,
in a completely covariant framework. It will not only reinforce the
claims made in the above work but also at the same time will make the
results independent of any slicing.

\subsection{Spherical Horizons}
\label{sec:sph_hor}
The plan is to use these considerations to find the toplogical
theory on the inner boundary called WIH. To proceed further, we consider
the following case: the fields on the boundary are such that the energy mommentum tensor is of the form $-T_{a}^{b}\l^{a}=e\l^{b}$, where $e$ is spherically symmetric. Then, using Einstein equation, we get:
\begin{equation}\label{sphere_phie_exp}
\Phi_{11}+\frac{1}{8}R -\frac{1}{2}\Lambda\=4\pi G e
\end{equation} 
This condition implies that for spacetimes with cosmological
constant zero, the term $\Phi_{11}+\frac{1}{8}R$ is spherically
symmetric as $e$ is spherically symmetric. From the above energy conditions with the above form of $-T_{a}^{b}\l^{a}$ it also follows simply from the Einstein
equations that $\Phi_{00}\=\Phi_{01}\=\Phi_{10}\=0$. This implies that no flux
of radiation falls through the horizon. This is expected as the
spherical symmetric energy condition is a special case of the more general
cases considered so far. 

We need one further expression to proceed further. First, see that for spherical symmetric horizons $\pi\=0$ and $\lambda\=0$ and $\mu$ is a real and spherically symmetric
function. In fact, in canonical formulation, this function
measures the expansion of the null vector $n$. In general black hole 
horizons, the function $\mu$ is positive but might not be spherically
symmetric. Indeed, for distorted black hole horizons, it is true that
$\mu$ is not spherically symmetric and $\pi$ is not zero. The Riemann
tensor is then calculated for the null vector $n^a$ by using (\ref{exp_normal_n}) as follows: 
\begin{eqnarray}
R_{abcd}~n^{d}=2\nabla_{[a}\nabla_{b]}n^{c}&\=&2\nabla_{[a}\omega_{b]}n_{c}+2\partial_{[a}\mu~\bar m_{b]}~m_{c} +2\mu~\nabla_{[a}\bar m_{b]}m_{c} \nn
&+& 2\mu~\bar m_{[b}\nabla_{a]}m_{c}+2\partial_{[a}\mu~ m_{b]}~\bar m_{c}
+2\mu~\nabla_{[a} m_{b]}\bar m_{c} \nn 
&+& 2\mu~ m_{[b}\nabla_{a]}\bar m_{c}+2\mu \omega_{[a}\bar m_{b]}m_{c}+ 2\mu \omega_{[a} m_{b]}\bar m_{c} 
\end{eqnarray}
We can use the expansion of the Riemann tensor in terms of the Weyl
tensor, the Ricci tensor and the Ricci scalar. Transvecting with $\l^a m^b \bar m^c$, and using the expansion of the scalar $\Phi_{11}$, we get that:
\begin{equation}\label{sphere_grad_n}
\mathbf{\Psi_2} +\frac{1}{12}R\=\lie_{\l}~\mu +\kappa_{(\l)}\mu
\end{equation}
This equation then implies that since the all the terms on the right
hand side are real and $R$ is real, the term $\mathrm{Im}\mathbf{\Psi_2}=0$,
\emph{i.e.}, $d\omega^{(\l)}=0$. Also, the equation
implies that the term $(~ \mathrm{Re}\mathbf{\Psi_2} +\frac{1}{12}R~)$ 
is again spherically symmetric as $\mu$ is spherically symmetric.

To proceed further, we want to restrict ourselves 
to the fixed area phase space. This needs an expression of the 
area two- form to be fixed on the phase space. We have already noticed that for spherical horizons, $d\omega^{(\l)}=0$.
However, the curvature of the one form $V^{(m)}$ involves the area form. The details of the
calculation is given in the appendix (see appendix \ref{sec:calc_dv}). Using the
equations (\ref{sphere_grad_n}) and (\ref{sphere_phie_exp}), we see
that the term $\mathcal{F}:=({\mathbf{\Psi_{2}}}^{(H)} -\Phi_{11} - \frac{R}{24})$ in the equation (\ref{exp_dv_epsilon}) is again 
spherically symmetric. This property of $\mathcal{F}$ remains
to be spherically symmetric throughout $\Delta$. To see this, note that
the  Bianchi identity for $V^{(H)}$ in equation (\ref{exp_dv_epsilon}) 
implies:
\begin{equation}
d\mathcal{F}\wedge~ ^2\mbf{\epsilon}\=0
\end{equation}
where, we have used that $ d~^2\mbf{\epsilon}\=0$. Transvected by $\l^a$,
the above equation 
shows that the value  of $({\mathbf{\Psi_{2}}}^{(H)} -\Phi_{11} - \frac{R}{24})$
is lie dragged by $\l^a$ and hence, remains fixed over $\Delta$. We want to
find a value for $\mathcal{F}$ in (\ref{exp_dv_epsilon}).  Then, we first evaluate the curvature of the connection $V^{(H)}=-im_{[I}\bm_{J]}A^{(H)}{}^{IJ}$ by using the Gauss- Bonnet theorem.
The connection $iV^{(m)}$ is precisely the connection on the sphere $S^2$ and then its
field strength will be the curvature.
Integrating both sides of (\ref{exp_dv_epsilon}), using Gauss- Bonnet theorem
and remembering that $\mathcal{F}$ is independent
of sphere coordinates, we get:  
\begin{equation}\label{exp-psi-2}
({\mathbf{\Psi_{2}}}^{(H)} -\Phi_{11} - \frac{R}{24})=-\frac{2\pi}{\mathcal{A}^{s}}
\end{equation}
This equation (\ref{exp-psi-2}) and the equation (\ref{exp_dv_epsilon}) together
imply that the two- form  $ ^2\mbf{\epsilon}$  can be written in terms
of the curvature of the $U(1)$ field $V^{(H)}$ :
\begin{equation}\label{epsilon_exp_sphere}
{}^2\mbf{\epsilon}=-\frac{\mathcal{A}^{s}_{\Delta}}{2\pi}~dV^{(H)}_{s}
\end{equation} 
The superscripts and subscripts $s$ in $\mathcal{A}^{s}_{\Delta}$  and
$V^{(H)}_{s}$ denote that we are in spherically
symmetric case.
As we shall see, this term will give the requisite level for the $U(1)$ Chern-
Simons theory. One can envisage another approach. Consider all spacetimes which
provide different set of values for $\left[{\mathbf{\Psi_{2}}}^{(H)} -\Phi_{11} -\frac{R}{24}\right]$.
Now, we shall consider only the average values. Then integrating the Eq.. (\ref{exp_dv_epsilon}),
we get an average value of $\left[{\mathbf{\Psi_{2}}}^{(H)} -\Phi_{11} -\frac{R}{24}\right]$ to be $\frac{-2\pi}{\mathcal{A}^{s}}$. While this is a possibility, We shall use (\ref{epsilon_exp_sphere})
in future. This is the most important expression in the derivation of
the $U(1)$ Chern -Simons theory. We shall use this expression for the
surface contribution to symplectic current for the case when the
the areas are fixed. This will imply that we will always remain in the
fixed area phase space. The symplectic structure will 
not be able to give any first law as the variation of the area is
zero on this phase space. We shall see however that the surface contribution
to the symplectic structure acquires the form of a $U(1)$ Chern- Simons theory.

To prove that claim, we go back to the expression for the symplectic current.
We have already seen that the symplectic current on the spacetime
region bounded by the Cauchy surfaces $M_{+},~M_{-}$ and $\Delta$ is given by: 
\begin{equation}
(\int_{M_{+}}- \int_{M_{-}})J(\delta_{1}, \delta_{2})=-\int_{\Delta}J(\delta_{1}, \delta_{2})
\end{equation}
In the expression for the symplectic current on $\Delta$, \emph{i.e.}
$\int_{\Delta}J(\delta_{1}, \delta_{2})$, the potentials $\psi_{(\l)}$ and
$\mu_{(m)}$ come into play(see (\ref{exp_j_delta_d})). The potential $\psi_{(\l)}$ is a function of $v$
only while $\mu_{(m)}$ is still a function of $(\theta, \phi)$. Then on
the fixed area phase space, the contribution to the symplectic current 
comes only from the terms involving the potential $\mu_{(m)}$.
The contribution to $\Delta$ for the spherical horizons can be calculated for
fixed area horizon. A simple calculation gives:
\begin{equation}
\int_{\Delta}J(\delta_{1}, \delta_{2})=\frac{1}{8\pi G \gamma}(\int_{S_{-}}- \int_{S_{+}})\{ \delta_{1}\mu_{(m)}~ \delta_{2}{}^2\mbf{\epsilon}
-(1 \leftrightarrow 2)\}
\end{equation}

 The one- forms $m$ have a gauge freedom. This is given by $m(v,\theta,\phi)=e^{-i\mu_{(m)} (v,\theta,\phi)} m(0,\theta, \phi)$. 
From now on, we shall only indicate the $v$ dependence of $\mu_{(m)}$. This gives that $V_{(m)}\rightarrow
V^{(m)}{}^{g}=V^{(m)}- i~d\mu_{(m)}$ (see (\ref{trans_rule_uuv})). The connection $V^{(H)}$ then transforms as
\begin{equation}\label{exp_HolstV_sphere}
V^{(H)}{}^{g}= V^{(H)} + \frac{1}{2}d\mu_{(m)}(v)
\end{equation}
To proceed further, we use the expression (\ref{epsilon_exp_sphere}) in the 
symplectic current and integrate by parts and again use (\ref{exp_HolstV_sphere}). This gives the following expression for the 
current
\begin{eqnarray}
\int_{\Delta} J(\delta_{1}, \delta_{2})&=& \frac{2}{8\pi G \gamma}\frac{\mathcal{A}^{s}_{\Delta}}{\pi}(\int_{S_{-}}- \int_{S_{+}})\{ \delta_{1}V^{(H)}{}^{g}~\wedge \delta_{2}V^{(H)}{}^{g}- (1 \leftrightarrow 2) \}\\ \nonumber
& -&\frac{1}{8\pi G \gamma}\frac{\mathcal{A}^{s}_{\Delta}}{\pi}(\int_{S_{-}}- \int_{S_{+}})\{ \delta_{1}V^{(H)}{}^{g}\wedge\delta_{2}V^{(H)}- (1 \leftrightarrow 2)\}
\end{eqnarray}

The further evaluation of the symplectic current is based on the idea
that the  expression for the connection $V^{(H)}$ is such  that it cannot be subject to
any variation as we move along the direction of $v$. This is because, this connection
has $V^{(H)}$ has only the dependence on $(\theta, \phi)$. However, the information of the 
$v$ dependence is carried by the field $\mu_{(m)}$. This implies that $ V^{(H)}{}^{g}$ has 
all the information of the $v$ dependence. In short, the $v$ dependence of 
$ V^{(H)}{}^{g}$ has been transferred to $\mu_{(m)}$ leaving $V^{(H)}$ only with the angular dependence.
Then we can reduce the expression for the symplectic current:
\begin{equation}
\int_{\Delta} J(\delta_{1}, \delta_{2}) =\frac{1}{8\pi G \gamma}\frac{\mathcal{A}^{s}_{\Delta}}{\pi}(\int_{S_{-}}- \int_{S_{+}})\{ \delta_{1}V^{(H)}{}^{g}\wedge \delta_{2}V^{(H)}{}^{g}\}
\end{equation}

In other words, we get , that the symplectic structure of the topological theory on WIH is precisely the
symplectic structure of Chern- Simons theory. 
We will  refer the theory on $\Delta$ to be a $U(1)$
Chern- Simons theory. The full symplectic structure for the
spherically symmetric phase space of black hole spacetimes is:
\begin{equation}
\Omega(\delta_{1}, \delta_{2})=\frac{1}{16\pi G\gamma}\int_{M}\left[ \delta_{1}(e^{I}\wedge
e^{J})~\wedge\delta_{2}A^{H}_{IJ} -\delta_{2}(e^{I}\wedge
e^{J})~\wedge\delta_{1}A^{H}_{IJ} \right] -\frac{1}{8\pi G \gamma}\frac{\mathcal{A}^{s}_{\Delta}}{\pi}\int_{S}\{ \delta_{1}V^{(H)}{}^{g}\wedge \delta_{2}V^{(H)}{}^{g}\}
\end{equation}

The level of the Chern- Simons theory is $-\frac{1}{8\pi G \gamma}\frac{\mathcal{A}^{s}_{\Delta}}{\pi}$. It is known that the level of the $U(1)$ Chern- Simons theory is an integer. So, we shall take $\frac{1}{8\pi G \gamma}\frac{\mathcal{A}^{s}_{\Delta}}{\pi}$ to be a positive integer. The result is highly non trivial considered in the backdrop of the WIH formlation. What we have shown  is that for
all spherical horizons, extremal or non- extremal, the topological theory is still the Chern- Simons theory. The Chern- Simons
gauge field does not see the $\xi$ scaling of the null normal $\l^a$, which controls the value of surface gravity
for the horizon. This simply implies that whatever be the null normal or whatever be the value of the surface gravity, the
effective symplectic structure on the horizon is still the Chern- Simons theory.

\section{Minimally Coupled Maxwell Fields on WIH}\label{emfield}
A horizon should also be able to hold matter fields on it. In this section, we
will give a general treatment of matter fields on WIH and find out the
constraints that might be placed on the matter fields. We will take the
example of electromagnetic field to analyse the situation. There will be some
degree of simplification for the use of Maxwell fields rather than any
arbitrary matter field,  but the treatment will be general in the sense that
the main results will be same for all other matter fields minimally coupled to
gravity. The reason we are interested in the Einstein-Maxwell system is
because our main objective is analysing the nonextremal and extremal horizons
through a unified formulation and the Einstein-Maxwell system provides the
finest set of examples of these types of horizons.  It is thus imperative to
check that the WIH boundary conditions have sufficient structure to enable the
existence of electromagnetic zeroth law and a first law.

This subsections are arranged in the following way. In the first, we will
recall the boundary conditions of WIH which will put restrictions on the
matter fields (Maxwell field), study its consequences and introduce conserved
charges defined on the cross-section of the WIH. We will in the process show
the main result that electromagnetic field can flow only along the horizon and
none can cross it. This is a general result and can be shown to be true for
all matter fields on $\Delta$. In the next subsection, we will prove that
zeroth law and the go on to prove the first law for the Einstein-Maxwell
system in the third subsection.

\subsection{Constraint on fields from boundary conditions}
Let us begin by recalling the boundary conditions of WIH which are of
importance here.  The  only way that WIH boundary conditions can restrict
matter is through conditions on the stress-energy tensor $T_{ab}$.  Thus,
constraints on matter fields will essentially come from the NEH boundary
condition since further restriction of NEH to WIH only restricts the class of
functions that can multiply the null normal $\ell^a$, so that the \emph{zeroth
law} is obeyed, and hence cannot put further restrictions on matter. Now, the
basic result that was obtained by the use of NEH boundary conditions and
Raychaudhuri equation is
\begin{equation}\label{Ricci_Contrac}
R_{ab}\l^a\l^b\=0
\end{equation}
It was also argued that this result holds for any null normal in the
equivalence class $[\xi\l^a]$.  We had already pointed out that this implies,
$R_{\ub{a}b}\l^b\=0$, Eq.. 
 for any null normal in
$[\xi\l^a]$. The consequences of these are
\begin{eqnarray}\label{NP_matterfield_values}
\Phi_{00}&=&\frac{1}{2}R_{ab}\l^a\l^b\=0 \nn \Phi_{01}&=&\frac{1}{2}R_{ab}\l^a
m^b\=0\nn \Phi_{10}&=&\frac{1}{2}R_{ab}\l^a \bar{m}^b\=0
\end{eqnarray}
These are the basic results and are true for all the matter fields on the
horizon and any null-normal in $[\xi\l^a]$. However, it is useful to study
these on a case by case basis for each of the matter fields since, as will be
shown below, the form of the Maxwell energy -momentum tensor introduces
further simplifications. To check those, let us notice that the Einstein field
equation and Eq.. (\ref{Ricci_Contrac}) implies that:
\begin{equation}
T_{ab}\l^a\l^b\=0
\end{equation}
for any $\l^a$ in $[\xi\l^a]$. An immediate consequence of this and
Eq..
is that
\begin{equation}\label{emtensor_cond}
T^{a}{}_{b}\l^b\=-e\ell^{a},
\end{equation}
for some non negative function $e$ on $\Delta$ and any $\l^a$ in $[\xi\l^a]$
. As we shall explicitly show and also is clear, this result physically
implies that there is no flux of radiation crossing the horizon, implying
isolation.

Let us now concentrate on the case of electromagnetic field. We will denote
the electromagnetic counterparts by bold letters. The main condition on the
field is thus
\begin{equation}\label{emenergycond}
\bfT_{ab}\ell^a \ell^b \triangleq 0
\end{equation}
for any $\l^a$ in $[\xi\l^a]$. The stress-energy tensor for electromagnetic
fields is given in terms of the field strength $\bfF=d\bfA$ as
\begin{equation}\label{emtensor}
{\bfT}_{ab} = \frac{1}{4\pi}[\bfF_{ac} {\bfF_b}^c - \frac{1}{4}g_{ab}
\bfF_{cd}\bfF^{cd}].
\end{equation}
Let us now argue what to expect. Contracting Eq..(\ref{emtensor}) with
$\ell^a$ on both the free indices, Eq. (\ref{emtensor_cond}) implies that the
vector $\ell^a \bfF_{ac}$ is null. Moreover, since $\bfF$ is antisymmetric,
the vector is  also normal to $\ell^a$. Thus, we can conclude that $\ell^a
\bfF_{ac}$ is proportional to $\ell_a$ and hence $\ell\lrcorner\ub{\bfF}\=0$.
This result will obviously be true for any $\l^a$ in $[\xi\l^a]$. To check
this result explicitly, we contract the expression of $\bfT_{ab}$ with $\ell^a
\ell^b$ for a fixed $\l^a$ in $[\xi\l^a]$ and check consequences for
$\bfF$. With Eq.. (\ref{emenergycond}) in mind, the first term in
Eq..(\ref{emtensor}) can be written as
\begin{eqnarray}
\bfF_{ac}\bfF_{db}\ell^{a} \ell^{b} g^{cd} &=&\bfF_{ac}\bfF_{db}\ell^{a}
\ell^{b}\left(m^{c} \bar m^{d} -\bar m^{c} m ^{d}\right) \nn &=&2
\left(\bfF_{ac}\ell^{a} m^{c} \right)\left(\bfF_{ac}\ell^{a} \bar m^{c} \right)
\end{eqnarray}
We have used the fact that the metric at the horizon can be expressed in terms
of a null-tetrad as $g_{ab}= -2\ell_{(a}n_{b)} +2m_{(a}\bar{m}_{b)}$ in the
first step and the anti-symmetry of $\bfF$ in the second. Similarly, it can be
checked that the second term in the Eq..(\ref{emtensor}) vanishes resulting in
\begin{equation}\label{emcons1}
0 \triangleq \bfT_{ab}\ell^a \ell^b \triangleq \mid\! \ell^a m^b \bfF_{ab}
\!\mid ^2 \, ,
\end{equation}
An immediate consequence of eqn. (\ref{emcons1}) is that
$\bfF_{ab}\ell^a=a\ell_{b}+b n_{b}$, where $a$ and $b$ are some arbitrary
functions. Contraction with $\ell^b$ and use of antisymmetry property of
$\bfF$ implies that $b=0$ and hence, we get
\begin{equation}\label{pbackf1}
\ub{\ell^a \bfF_{ab}} \triangleq 0.
\end{equation}
for any $\l^a$ in $[\xi\l^a]$. In order to  obtain a similar expression for
$\sbfF$ recall that the stress energy tensor can be written as
\begin{equation}\label{dualemtensor}
\bfT_{ab} = -\frac{1}{4\pi} [\sbfF_{ac} {\sbfF_b}^c - \frac{1}{4}g_{ab}
\sbfF_{cd}\sbfF^{cd}]
\end{equation}
Using arguments which led to eqn. (\ref{pbackf1}), we obtain a similar
restriction on $\sbfF$:
\begin{equation}\label{dualpback}
\ub{\ell^a  {\sbfF}_{ab}} \triangleq 0.
\end{equation}
It is straightforward to show that eqn. (\ref{pbackf1}) and
eqn. (\ref{dualpback}) puts further constraints on the electromagnetic field
tensor. To observe this, note that one can write $F_{ab}\ell^{a}\triangleq
a\ell_b$, for any $\l^a$ in $[\xi\l^a]$ and a similar one for the dual
$\sbfF$. Then using the expressions (\ref{emtensor}) and (\ref{dualemtensor}),
the following conditions can be easily checked.
\begin{eqnarray}
\bfT_{ab} \ell^a m^b &\triangleq& 0 ~ \triangleq \bfT_{ab} \ell^a \bar{m}^b\nn
 \bfT_{ab} m^a m^b &\triangleq& 0 ~\triangleq \bfT_{ab} \bar{m}^a \bar{m}^b.
\end{eqnarray}

It is interesting to observe that the first two set of conditions are none
other than the ones we already had obtained in
eqn. (\ref{NP_matterfield_values}) and thus are universal for any matter field
on WIH. The second set of conditions however are special for the
electromagnetic fields. In terms of Newman-Penrose components, these imply the
following restrictions on the Ricci tensor:
\begin{eqnarray}\label{NP_comp_emspecial}
\Phi_{02}&:=&\frac{1}{2}R_{ab}m^a m^b\=0\nn
\Phi_{20}&:=&\frac{1}{2}R_{ab}\bar{m}^a\bar{m}^b\=0
\end{eqnarray}

Now, we can make some statements about the \emph{isolation} of the WIH. Given
a global timelike Killing vector field $\tau^a$, the Poynting vector,
describing the direction of energy flow, is defined as $T^a{}_b \tau^b$. On a
WIH, the corresponding term can be defined as $T^a{}_b \ell^b$ which by the
boundary conditions is future directed and causal (null).  Using
eqn. (\ref{emenergycond}) and the first set in eqn. (\ref{NP_comp_emspecial}),
we can safely say that $T^a{}_b \ell^b$ is proportional to $\ell^a$. In a
local coordinate system adapted to WIH, $\ell^a=\left(
\frac{\partial}{\partial v}\right)^a $.  Thus the direction of the energy flow
is along the horizon and nothing can cross the WIH though there is no
restriction on the presence of radiation even arbitrary close to the
horizon. For e.g., in a local advanced Eddington -Finkelstein coordinates $(v,
r, \theta, \phi)$, it is easy to show that the components $F_{rv},
F_{r\theta},F_{r\phi} $ may exist close to the horizon and are unrestricted in
the values but have no contribution when restricted and pulled back to the
horizon.

Let us now define the electric and magnetic flux density two forms directed
outwards.  The electric flux two form is given by ${\bf E}_{\Delta} \= -\sbfF$
and the magnetic one by ${\bf B}_{\Delta} \= -\bfF$.  The signatures have been
taken such so as to take the orientation of $S_{\Delta}$ ( $S_{\Delta}$ is a
cross-section of $\Delta$, see fig. 1) into account which is defined with
respect to normal pointing into the horizon.  Let us first evaluate
\begin{equation}
\lie_{\xi\ell}\ub{\bfF} \triangleq \xi \ell \cdot \ub{d\bfF} + \ub{d(\xi\ell
    \cdot \bfF)}.
\end{equation}

The first term on the right hand side vanishes due to Maxwell's equations on
$\Delta$, while the second term is zero due to the previous restriction on
$\bfF$, eqn. (\ref{pbackf1}). Therefore we conclude that $\ub{\bfF}$ is Lie
dragged by any $\ell^a$ in $[\xi\l^a]$.  An identical argument for $\sbfF$
leads to the analogous conclusion.  Therefore we obtain
\begin{equation}\label{lieemF}
\lie_{\xi\ell}\ub{\bfF} \triangleq 0 \qquad \mbox{and} \qquad
\lie_{\xi\ell}\ub{\sbfF} \triangleq 0 \, .
\end{equation}
These results imply that the $2-$forms ${\bf E}_{\Delta}$ and ${\bf B}_{\Delta}$ are 
``time-independent". However, these do not restrict the forms
of these fields otherwise.

We can now define the electric charge of the horizon (we assume that the
magnetic charges are zero, which if present, can be analogously defined).
Since the horizon is an inner boundary of spacetime, the normal to a 2-sphere
cross section of the horizon will naturally be inward pointing. Bearing this
in mind, we define the electric charge of the horizon as
\begin{equation}\label{charge}
Q_{\Delta} :\triangleq-\frac{1}{4\pi} \oint_{S_{\Delta}} \sbfF
\end{equation}
For the definition to be meaningful,we should ensure that the values of
$Q_{\Delta}$ should be independent of the cross section of the horizon
$S_{\Delta}$. This result can be anticipated since the NEH boundary conditions
imply that a Killing vector field exists on $\Delta$, one expects that the
charge to be independent of cross-section.  Since the $2-$ forms ${\bf
E}_{\Delta}$ and ${\bf B}_{\Delta}$ are ``time-independent", this guarantees
that $Q_{\Delta}$ is independent of the choice of cross section $S_{\Delta}$
of the horizon.  Note that this result was obtained using only the boundary
conditions; equations of motion in the bulk are not needed.

\subsection{Electromagnetic Zeroth Law}
The zeroth law for electromagnetic field states that one can define a scalar
potential on the horizon that is constant throughout the horizon. Thus, to
establish the zeroth law for the electromagnetic case, we need to define an
electric potential $\Phi$ at the horizon. For this, the electromagnetic
potential ${\bf A}$ is gauge fixed on $\Delta$ such that
\begin{equation}\label{lieema}
\lie_{(\xi\ell\,)}{\bf A}_{\ub a}\triangleq\grad_{\ub a}\lie_{(\xi
\ell\,)}\chi_{(\xi\ell\,)}\;.
\end{equation}
where $\chi_{(\xi\ell\,)}$ is arbitrary non-zero but a fixed function of $v$
alone. Following eqn. (\ref{lieema}), given such an electromagnetic potential
${\bf A}$ we can now define the scalar potential $\Phi_{(\xi\ell\,)}$ at the
horizon as
\begin{equation}\label{emscalar_pot}
\Phi_{(\xi\ell\,)}\triangleq
-\xi\ell\cdot{\bfA}+\lie_{(\xi\ell\,)}\chi_{(\xi\ell\,)}
\end{equation}

In a flat spacetime, the scalar potential is defined as the time component of
the gauge potential one form $\bfA$. However, this requires a gauge fixing
since $\bfA$ takes values in the gauge equivalence class of addition of an
exact form, $\bfA\rightarrow \bfA + d\lambda$, $\lambda$ being some arbitrary
function. So the scalar function also suffers from a gauge ambiguity of adding
a total time derivative.

Recall that the Einstein-Maxwell case (static space-times) involves the
electro-static potential $\Phi$ which one typically sets $\Phi = -
\tau^a\bfA_a$ where $\tau^a$ is the static Killing field and the gauge is
chosen such that the vector potential $\bfA$ tends to zero at infinity and
satisfies ${\cal L}_\tau \bfA =0$ \textit{everywhere in space-time}. Note that
under the electromagnetic gauge transformation $\delta_{\lambda} \bfA_{a}=
\nabla_{a}\lambda$, the eqn. (\ref{lieema}) reduces to
\begin{equation}\label{reducedemA}
\lie_{(\xi\ell\,)}{\bf \bar A}_{\ub a}\triangleq\lie_{(\xi \ell\,)}\grad_{\ub
a}\left[ \chi_{(\xi\ell)} -\lambda\right]
\end{equation}
If we gauge fix $\chi_{(\xi\ell)}=\lambda$, the definition used in static
spacetime is obtained.

Thus the definition we are proposing here is completely consistent and more
general than is usually used. Then eqn. (\ref{emscalar_pot}) is just the
standard definition of scalar potential, $\xi\l^a$ playing the role of
''time''  on $\Delta$ and the additional term in eqn. (\ref{emscalar_pot}) is
just a total time derivative. It follows immediately that
\begin{equation}
d\Phi_{(\xi\ell\,)}\triangleq 0
\end{equation}
hence $\Phi_{(\xi\ell\,)}$ is constant on the horizon which essentially is the
\emph{electromagnetic zeroth law}. In order that this is true for the entire
equivalence class $[\,\xi\ell^a\,]$ requires the gauge fixing functions to
vary in the class in a specific way,  $\lie_{(\xi
\ell)}\chi_{(\xi\ell\,)}-\Phi_{(\xi\ell\,)}
=\xi[\lie_{\ell}\chi_{(\ell\,)}-\Phi_{(\ell\,)}]$. This restriction is to be
viewed as follows: it is always possible to choose $\chi_{(\ell)}=0$ for one
$\ell^a$ such that $\Phi_{(\ell)}$ is a constant on $\Delta$.  Then for each
null vector $\xi\ell^a$, the above restriction fixes the gauge in
eqn. (\ref{emscalar_pot}) such that $\Phi_{(\xi\ell)}$ remains a constant on
$\Delta$. For constant rescaling of $\ell^a$, it is consistent to choose
$\chi_{(\ell)}=0$ for all $\ell^a $ (like one does for the flat spacetime) but
is not true for the generalized class $[\xi\ell^a]$. It is a nontrivial fact
that even for the generalized class $[\xi\ell^a]$, a constant potential such
as in eqn. (\ref{emscalar_pot}) exists making use of the gauge ambiguity
(which always exists for scalar potentials) and the boundary conditions alone.

\subsection{Electromagnetic First Law}
The electromagnetic part of the Lagrangian four-form 
is given by $8\pi L=-{\bf F}\wedge*{\bf F}$. The variation of this Lagrangian is
carried out over all ${\bf A}$s that have the expected asymptotic fall-offs and are
gauge fixed on $\Delta$ as in (\ref{lieema}). The key point to note is that although
a surface term is needed in the gravitational part of the action, thanks to the
electromagnetic zeroth law, such a term is not needed for the electromagnetic part.
Proceeding as before
we find a bulk and a surface symplectic structure. To extract the surface
term we introduce a potential for $\Phi_{(\xi\ell\,)}$ (just like what we did for
$\kappa_{(\xi\ell\,)}$): $\lie_{(\xi\ell\,)}\varphi_{(\xi\ell\,)}\triangleq-\Phi_{
(\xi\ell\,)}$. It also
suffers from an additive ambiguity which is removed by choosing
$\varphi_{(\xi\ell\,)}|_{S_-} \triangleq 0$. Then the electromagnetic part of the symplectic structure becomes
\begin{eqnarray}\label{totalsymplectic}
 &&\Omega_{\rm em}(\delta_1,\delta_2)=
-\frac{1}{4\pi}\!\int_M\!\big[\delta_1\!*{\bf F}\wedge\delta_2{\bf A}
       -(1\leftrightarrow 2)\big]+\nn &&
 \frac{1}{4\pi}\oint_{S_\Delta}\!\big[\delta_1\!*{\bf F}\,\delta_2
 (\chi_{(\xi\ell\,)}+\varphi_{(\xi\ell\,)})-(1\leftrightarrow 2)\big]\;.
\end{eqnarray}

Again, we wish to evaluate $X_{(\xi\ell\,)}$ from the electromagnetic part of the
symplectic structure. Making use of the field equations we find that the bulk symplectic
structure gets contributions only through the boundaries, which equals
$(\lie_{(\xi\ell\,)}\chi_{(\xi\ell\,)}-\Phi_{(\xi\ell\,)})\,\delta Q_\Delta$. To
evaluate the
contribution from the surface symplectic structure care should be taken not to
equate $\delta_{(\xi\ell\,)}$ with $\lie_{(\xi\ell\,)}$ for the potential $\varphi$.
It turns out that $\delta_{(\xi\ell\,)}\varphi_{(\xi\ell\,)}\triangleq 0$ everywhere 
and the contribution is $-\delta_{(\xi\ell\,)}\chi_{(\xi\ell\,)}$. Combining
contributions from the bulk and the surface, we find
\begin{eqnarray}\label{oneformtotal}
X_{(\xi\ell\,)}^{\rm em}(\delta)\triangleq-\Phi_{(\xi\ell\,)}\delta Q_\Delta\;.
\end{eqnarray}
Thus, the combined first law for the gravitational and electromagnetic fields agrees
with the standard first law of black hole thermodynamics.

\subsection{E.M. Contribution to Surface Symplectic Structure}
We have already obtained that $\ub{\bfF_{ab}\l^a}\=0$. This implies that:
\begin{equation}\label{form_of_em_F}
\bfF_{ab}\=\bar{\alpha}_{[b}m_{a]}+{\alpha}_{[b}\bm_{a]}+\beta_{[b}\l_{a]}
\end{equation}
where, $\alpha$ is a complex one form and $\beta$ is a real one form such 
that $\bfF_{ab}$ is real. However, it is also imperative that one needs
more conditions on the one forms $\alpha$ and $\beta$ to match the degree of
freedom  of the $\bfF$ tensor.The conditions are also obtained from the
previous condition eqn (\ref{form_of_em_F}). They are:
\begin{eqnarray}
\alpha_{a}\l^a\=&0&\=\bar{\alpha}_{a}\l^a\ \nn
\beta_{a}\l^a&\=&0
\end{eqnarray}

This imples that $\alpha_{a}=a m_{a} +b \bm_{a}$ and  $\beta_{a}=\bar{c} m_{a}
+c \bm_{a}$. Then, we write the expansion of $\bfF$ in terms of the geometric 
forms $(\l, n, m, \bm)$. Then, we obtain that:
\begin{equation}\label{exp_em_F_mmbar}
\ub{\bfF_{ab}}\=\bar{a}~ \bm_{[b}m_{a]} + a~ m_{[b}\bm_{a]}
\end{equation}
Now, note that in NP formalism, the six components of $\bfF$ are expressed
in terms of the three complex scalars $\phi_{0}, ~\phi_{1}$ and  $\phi_{2}$.
It can be observed that $\phi_{0}\=0$. The other scalar is
$\phi_{1}=\frac{1}{2}\bfF_{ab}\left(m^{a} \bm^{b}-\l^{a} n^{b}\right)$.
We can express the above equation eqn.(\ref{exp_em_F_mmbar})
in terms of the complex scalars. This gives
$(a-\bar{a})=(\phi_{1}-\bar{\phi}_{1})$. Then, the expansion is :
\begin{equation}
\ub{\bfF_{ab}}\=-(2~ Im\phi_{1}){}^{2}\epsilon
\end{equation}

 Similarly, it can be argued that the expansion of $\sbfF$ is :
\begin{equation}
\ub{\sbfF_{ab}}\=(2~ Re\phi_{1}){}^{2}\epsilon
\end{equation}

We now argue that on the phase space of fixed parameters, the
symplectic structure of Maxwell fields does not contain any
surface term. To prove this, we first see that the
symplectic current  is given by:
\begin{equation}
J(\delta_{1}, \delta_2 )=\frac{1}{4\pi}\left(\delta_1 \sbfF\wedge \delta_2 A - \delta_2 \sbfF\wedge \delta_1 A  \right)
\end{equation}
The next crucial step that one should take is to check wheather the integral
of the symplectic current, integrated over $\Delta$ (as in fig. 1)
is zero on the space of solutions. In general, this is not true.
However, when we consider those histories where the parameters are held fixed,
the symplectic current goes to zero when integrated over $\Delta$.This
means that on the parameter-fixed phase space, the Maxwell theory
does not contribute to the surface symplectic structure.

\section{Discussions}
The Weak Isolated Horizon boundary conditions had been shown to be weak enough
to include the extremal and non-extremal horizons in the same phase space \cite{cg}.
This extension opens up the possibility of an understanding of entropy of extremal black holes in
supergravity and string theory.
Extremal black holes play a fundamental role in supergravity and the string theories (see \cite{bdewit} for discussions
and other references).
These solutions possess a high degree of
supersymmetry as isometries and due to some non-renormalization theorems one expects
the counting of degeneracies of the associated quantum state to be protected over a range of 
string coupling constant which can vary from small to large values. Thus, the results obtained 
perturbatively, at small values of the parameter, continue to hold for other large values.
The popular choice for calculating the entropy of such black holes is the Killing horizon (KH)
framework and the classical \emph{entropy formula}
suggested in \cite{Wald}. Inspite of its wide use, it is to our opinion highly unsatisfactory:
Firstly, the derivation requires the existence of bifurcation two-spheres
and it is not clear how to generalise the
framework to extremal black holes which admit no bifurcation two-sphere.
In other words, the phase-space of non-extremal black holes does not include extremal solutions.
Secondly, the extremal solutions arising in such theories
have curvature singularities at the horizon when the extremal limit is taken. This
necessarily asks for higher order stringy corrections to be taken into account \cite{sen}. 
To find the entropy of these extremal black holes in string theory, it is argued that entropy
of these solutions
are to be defined {\em only through limits} from their non-extremal
counterparts (see \cite{gm} for the arguments).
However, since the phase-space of non-extremal Killing horizons do not contain extremal horizons, it becomes
ambiguous how such limits are to be taken.
The original formulation of isolated horizons (IH) \cite{afk}
bypass some of these difficulties but still it is not enough.
The extremal and non-extremal black holes continue to
remain in different phase spaces so far as the validity of the
first law is concerned and hence, \emph{extremal limit} doesn't make much sense.
A new framework, called \emph{weak isolated horizons} (WIH),
presented in \cite{cg}, removes this difficulty allowing one to take limits in the same phase space. 
Since in WIH, the extremal horizons are in the same class of the non-extremal ones, the
entropy of an extremal black hole is automatically determined when one quantizes this WIH.
Expectedly, the result is proportional to the area of the horizon.
This is consistent with the claim made in \cite{gm} that the entropy of an extremal black hole
must be proportional to the area of the horizon, provided one uses a phase space that contains
both extremal and non-extremal global solutions. A recent paper \cite{cjr} revisits some
of these arguments. The basic idea behind this paper was to check wheather
such expectations are borne out in a completely covariant manner.

In this paper we extended the formulation of Weak Isolated Horizons to Holst's action. The main reason for taking up this exercise is to make WIHs applicable
in the framework of loop quantum gravity which makes essential use of the Holst action (See \cite{fffp} for other applications of Holst's action). Through our analysis we show that the essential structures of the phase
space remain unaltered from the Palatini phase space once we rewrite everything
in terms of the Holst connection one-form $A^{(H)}$. Interestingly, there still
exists a boundary symplectic structure, although new potentials are needed here.
Both the bulk and boundary symplectic structures conspire in such a way that the
first law of WIH mechanics still holds. Although the results are very similar in
spirit with the ones obtained from the Palatini action, the two cases differ substantially in details which
have been elaborated in this paper. Another hallmark of our approach is that compared to the earlier results which used canonical phase space our approach is completely covariant. As a by-product we find that the effective theory at the horizon is a $U(1)$ Chern-Simons theory which is obtained here from a completely covariant framework (this is to our knowledge has not been derived earlier). We also argued that the presence of electromagnetic fields on WIH does not affect the boundary symplectic structure
and hence does not have any affect on the topological theory on the boundary.

\section{Acknowledgments}

We gratefully acknowledge discussions with Parthasarathi Majumdar and Parthasarathi Mitra in our group meetings. We especially thank the former for continuous encouragement and many stimulating remarks that
helped to improve our work. 

\section{Appendices}
\subsection{The Newman-Penrose formalism}
\label{appb}

In this part of the appendix, we give a brief summary 
of the  Newman- Penrose (NP) formalism (\cite{NP}). The details
of the formalism, notation and other technicalities are given
in (\cite{Chandra,Stewart}). Though we will follow
\cite{Chandra} so far as the notation is concerned, the signature
$(-,+,+,+),$ is different in our case (and is same as \cite{afk,cg}) and hence one
needs to recalculate some of the results.

The NP formalism relies on the fact that one is allowed to have a basis
with set of $4$ null
vectors ($\ell$, $n$, $m$, $\bar{m}$). The pair $\ell$ and $n$  are real while
pair $m$ and $\bm$ are complex conjugates of each other. These satisfy the
following \emph{orthonormality} conditions:
\begin{equation}
\l.n = -1 \qquad m.\bm = 1,
\end{equation}
the rest being equal to zero.
The next step is to define the Newman-Penrose spin coefficients
(also called the Ricci rotation coefficients in the tetrad formalism).
All the information that the connection provides is encoded in the
$12$ independent complex scalars. These are designated by special symbols
as is given below:
\begin{eqnarray*}&&
\kappa:= -m^a\ell^{b}\nabla_{b}\l_{a}=-m^{a} D\l_{a} \qquad \rho :=-m^{a}\bar{m}^{b}\nabla_{b}m_{a}=-m^{a}\bdelta m_{a}\\
&&\sigma :=- m^{a}m^{b}\nabla_{b}\ell_{a}=- m^{a}\delta\ell_{a} \qquad \mu :=-n^{a}m^{b}\nabla_{b}\bm_{a}=-n^{a}\delta\bm_{a}\\
&&\lambda :=-n^{a}\bar{m}^{b}\nabla_{b}\bm_{a}=-n^{a}\bdelta\bm_{a} \qquad \tau :=-m^{a}n^{b}\nabla_{b}\ell_{a}=-m^{a}\Delta\ell_{a}\\
&&\nu :=-n^{a}n^{b}\nabla_{b}\bm_{a}=-n^{a}\Delta\bm_{a} \qquad \pi := -n^{a}\l^{b}\nabla_{b}\bm_{a}= -n^{a}D\bm_{a}\\
&&\epsilon :=-\frac{1}{2}(n^{a}\ell^{b}\nabla_{b}\l_{a}+
m^{a}\l^{b}\nabla_{b}\bm_{a})=-\frac{1}{2}(n^{a}D\l_{a}+ m^{a}D\bm_{a}) \\
&&\gamma :=-\frac{1}{2}(n^{a}n^{b}\nabla_{b}\l_{a}+ m^{a}n^{b}\nabla_{b}\bm_{a})
=-\frac{1}{2}(n^{a}\Delta\l_{a}+ m^{a}\Delta\bm_{a}) \\
&& \alpha :=-\frac{1}{2}(n^{a}\bar{m}^{b}\nabla_{b}\ell_{a}
+m^{a}\bar{m}^{b}\nabla_{b}\bm_{a})=-\frac{1}{2}(n^{a}\bdelta\ell_{a}
+m^{a}\bdelta\bm_{a})\\
&&\beta :=-\frac{1}{2}(n^{a}m^{b}\nabla_{b}\ell_{a}
+m^{a}m^{b}\nabla_{b}\bm_{a})
=-\frac{1}{2}(n^{a}\delta\ell_{a} +m^{a}\delta\bm_{a}) .
\end{eqnarray*}
where, the symbols $D, \Delta, \delta, \bdelta$ are the directional
derivatives along the basis vectors $\l, n, m, \bm$, \emph{i.e.}
\begin{equation}
D = \ell^{a} \nabla_{a} \qquad \Delta = n^{a} \nabla_{a} \qquad
\delta = m^{a} \nabla_{a} \qquad \bar{\delta} = \bar{m}^{a}
\nabla_{a}.
\end{equation}
In other words, the derivative operator written in terms of these symbols turn
out to be: 
\begin{equation*}
\nabla_{a}=-n_{a}D-\l_{a}\Delta +\bm_{a}\delta+ m_{a}\bdelta.
\end{equation*}
The ten independent components of the Weyl tensor are expressed in
terms of five complex scalars $\Psi_{0}$, $\Psi_{1}$, $\Psi_{2}$,
$\Psi_{3}$ and $\Psi_{4}$. The ten components of the Ricci tensor are
defined in terms of four real and three complex scalars $\Phi_{00}$,
$\Phi_{11}$, $\Phi_{22}$, $\Lambda$, $\Phi_{10}$, $\Phi_{20}$ and
$\Phi_{21}$ . These scalars are defined as follows:
\begin{eqnarray*}&&
\Psi_{0} = C_{abcd}\ell^{a}m^{b}\ell^{c}m^{d} \qquad \Phi_{01} =\frac{1}{2}R_{ab}\ell^{a}m^{b} \qquad \Phi_{10} =\frac{1}{2}R_{ab}\ell^{a}\bar{m}^{b}  \\
&& \Psi_{1} =C_{abcd}\ell^{a}m^{b}\ell^{c}n^{d} \qquad \Phi_{02} =\frac{1}{2}R_{ab}m^{a}m^{b} \qquad \Phi_{20} =\frac{1}{2}R_{ab}\bar{m}^{a}\bar{m}^{b}  \\
&&\Psi_{2} = C_{abcd}\ell^{a}m^{b}\bar{m}^{c}n^{d}  \qquad  \Phi_{21} =\frac{1}{2}R_{ab}\bar{m}^{a}n^{b}  \qquad \Phi_{12} =\frac{1}{2}R_{ab}m^{a}n^{b} \\
   && \Psi_{3} =C_{abcd}\ell^{a}n^{b}\bar{m}^{c}n^{d}  \qquad
    \Phi_{00} =\frac{1}{2}R_{ab}\ell^{a}\ell^{b}  \qquad
    \Phi_{11} =\frac{1}{4}R_{ab}(\ell^{a}n^{b} + m^{a}\bar{m}^{b}) \\
   && \Psi_{4} =C_{abcd}\bar{m}^{a}n^{b}\bar{m}^{c}n^{d}  \qquad
    \Phi_{22} =\frac{1}{2}R_{ab}n^{a}n^{b}   \qquad
    \Lambda = \frac{R}{24}
\end{eqnarray*}
To express the Weyl tensor in terms of the five complex scalars, we proceed 
as follows \cite{Chandra}. First, we construct a product of four quantities
such that it has all the symmetries of the Weyl tensor. It should be
antisymmetric
in the first two indices and the last two indices and should remain unchanged 
under simultaneous interchange of the first two and the last two indices.  We denote that by
$\{\}$. For example,
\begin{eqnarray}
\{A_a B_b C_c D_d\}&:=&A_a B_b C_c D_d - B_a A_b C_c D_d - A_a B_b D_c C_d +
B_a A_b D_c C_d \nn &+& C_a D_b A_c B_d - D_a C_b A_c B_d -
C_a D_b B_c A_d + D_a C_b B_c A_d  
\end{eqnarray}
There is subtlety in this construction of $\{\}$. In case all the terms
in the braces are different from each other, just as in the 
above example all the terms are distinct, the  $\{\}$ is of the form as given above. 
In case the term is like
$\{A_a B_b A_c B_d\}$, having some indistinguishable terms, only the first set of four terms as given
above will suffice and it
will satisfy the criteria of being antisymmetric
in the first two indices and the last two indices and should remain unchanged 
under simultaneous interchange of the first two and the last two indices.
It is then trivial to check that the components of the Weyl tensor $C_{abcd}$ 
can be expanded as 
\begin{eqnarray}\label{weylexp}
C_{abcd}&=&F_{1212}\{\l_a n_b \l_c n_d\}+ F_{3434}\{m_a \bm_b m_c \bm_d\}+
F_{1234} \{l_a n_b m_c \bm_d\}\nn &+& F_{1314}\{\l_a m_b \l_c \bm_d\} +
F_{2324}\{n_a m_b n_c \bm_d\}\nn
&+& [F_{1313} \{\l_a m_b \l_c m_d \} + F_{2323} \{n_a m_b n_c m_d\} +
F_{1213}\{\l_a n_b \l_c m_d\}\nn
&+& F_{1223} \{\l_a n_b n_c m_d\} + F_{1323} \{\l_a m_b n_c m_d\} +  F_{1324}
\{\l_a m_b n_c \bm_d\}\nn
&+& F_{1334} \{\l_a m_b m_c \bm_d\}+ F_{2334} \{n_a m_b m_c \bm_d\} + ~~c.c. ]
\end{eqnarray}
where, $F_{1234}, \cdots$ are expansion coefficients to be determined. The
$c.c$ refers to the comples conjugates. Note that the terms outside the square
braces are manifestly real whereas the terms inside are complex. The complex
conjugates can be written down by simultaneous interchange of $m$ to $\bm$
and index number $3$ to $4$.
 
We now determine the terms in the expansion of $C_{abcd}$ $i.e.$ $F_{1234}, \cdots$. It is easy to check that: 
$F_{2424}=\Psi_0$, $F_{1224}=\Psi_1$, $F_{1324}=-\Psi_2$, $F_{1213}=-\Psi_3$ 
and $F_{1313}=\Psi_4$. The determination of other terms in the expansion of
$C_{abcd}$  requires some other relations. Firstly, cyclicity requires that:
\begin{equation}\label{cyclic}
C_{1234}+C_{1342}+C_{1423}=0
\end{equation}
The notation used here is the following: we mark $l=1,\; n=2,\; m=3,\; \bm=4$
so that whatever index is contracted gets the above assigned value, \emph{for e.g.}, $C_{1234}=C_{abcd}\l^{a}n^{b}m^{c}\bm^{d}$.
Now, since $C_{abcd}$ is trace free, it implies the following results. Firstly,
\begin{equation}\label{trace}
C_{1314}=C_{1332}=C_{2324}=C_{2441}=0
\end{equation}
and secondly, using eqn.(\ref{cyclic}) we get:
\begin{eqnarray}\label{trace1}&&
C_{1232}=C_{3234}\qquad C_{1231}=C_{1334}\qquad C_{1241}=C_{1443}\qquad
C_{1242}=C_{2434}\nn
&&C_{1212}=C_{3434}\qquad C_{1342}=\frac{1}{2}(C_{1212}-C_{1234})
\end{eqnarray}

Now, we will use these relations. First, note  that
$F_{1224}=-C_{2113}=-C_{1334}$. Thus, $-C_{1334}=F_{2443}=\Psi_1$.
Secondly, note that $\Psi^{\ast}_{2}=C_{1432}=-F_{1423}$. Using
eqn. (\ref{cyclic}), $C_{1234}=-(\Psi_{2}-\Psi^{\ast}_{2})$. Thus,
$F_{1234}=C_{2143}=-(\Psi_{2}-\Psi^{\ast}_{2})$. Also, using the 
last two equation in eqn. (\ref{trace1}), we get, $F_{1212}=C_{2121}=(\Psi_{2}+\Psi^{\ast}_{2})$
and $F_{3434}=C_{4343}==(\Psi_{2}+\Psi^{\ast}_{2})$. The rest,
$F_{1314}=F_{2324}=F_{1323}=F_{1424}=0$, follows from the eqn. (\ref{trace}).

Putting all these values of $F_{abcd}$ in the eqn. (\ref{weylexp}), we get
\begin{eqnarray}
C_{abcd}&=&(\Psi_{2}+\Psi^{\ast}_{2})\left[\{\l_a n_b \l_c n_d\}+\{m_a \bm_b m_c
  \bm_d\}\right]-(\Psi_{2}-\Psi^{\ast}_{2})\{ \l_a n_b m_c \bm_d\}\nn
&+&\left[\Psi_{4}\{\l_a m_b \l_c m_d\}+ \Psi_{0}\{n_a \bm_b
  n_c \bm_d\} -\Psi_{2}\{\l_a m_b n_c \bm_d\}+  complex ~conjugates\right]\nn 
&+&\left[\Psi_{1}\left(\{\l_a n_b n_c\bm_d\} +\{n_a \bm_b \bm_c m_d\}\right)+complex
  ~conjugates \right]\nn
&+&\left[\Psi_{3}\left(\{\l_a m_b m_c \bm_d\} -\{\l_a n_b \l_c n_d\}\right) + complex ~conjugates\right] 
\end{eqnarray}
These has been used used in finding the expression for $d\omega^{(\l)}$ 
and $dV^{(m)}$.

\subsection{Equivalence of Palatini and Holst Symplectic Structure}
\label{sec:Holst_Vs_Palatini}

 Consider the case when the manifold has no
boundary.  Then, the symplectic structure is given by:
\begin{equation} \Omega\left( \delta_1, \delta_2
\right):=\dfrac{1}{16\pi G\gamma}\int_{M}\left\lbrace \delta_{[1}\left( e_{1}\wedge
e_{2}\right) \right\rbrace \wedge\left\lbrace \delta_{2]}\left(
A_{IJ}-\frac{\gamma}{2} \epsilon_{IJ}{}^{KL}A_{KL}\right)\right\rbrace
\end{equation}
Now, the crucial point to note is the following. When viewed from the perspective of
phase-space, the Holst term is  a canonical transformation
on the phase space. We shall show that $\gamma$ dependent term will vanish.
even in presence of inner boundaries. 
We then must show that:
\begin{eqnarray}\label{HP1}  \delta_{1}\left( e^{I} \wedge
e^{J}\right)\wedge\delta_{2}A_{IJ}&=&\delta_{2}\left( e^{I} \wedge
e^{J}\right)\wedge\delta_{1}A_{IJ}\nn \Rightarrow \delta_{1}\left(
e^{I}\right)  \wedge \left(
e^{J}\wedge\delta_{2}A_{IJ}\right)&=&\delta_{2}\left( e^{I}\right)  \wedge
\left( e^{J}\wedge\delta_{1}A_{IJ}\right)
\end{eqnarray}

To prove the equality, we use the equation of motion. Firstly, from the
equation of motion, we get that
\begin{equation}\label{HP2}  e^{J}\wedge \delta_2 A_{IJ}=d\delta_2 e^I +
A_{IJ}\delta_2 e^J
\end{equation}
Putting this equation eqn. (\ref{HP2}) in the L.H.S. of (\ref{HP1}), we get,
\begin{equation}\label{HP3}  \delta_{1}\left( e^{I}\right)  \wedge \left(
e^{J}\wedge\delta_{2}A_{IJ}\right) =\delta_{1}\left( e^{I}\right)  \wedge
d\left( \delta_2 e_{I}\right)-A_{IJ}\wedge e^I\wedge\delta_2e^J.
\end{equation}
Also, the first term in (\ref{HP3}) can be further reduced as:
\begin{equation}\label{HP4}  \delta_{1}e^I\wedge d\delta_{2}e_{I}=-d\left(
\delta_1 e^I \wedge \delta_2 e_{I} \right)+d\delta_{1}e^I \wedge\delta_{2}e_{I}
\end{equation}

Thus, using the equation eqn. (\ref{HP4}), the equation (\ref{HP3}) becomes:
\begin{eqnarray} L.H.S.&=&-d\left( \delta_1 e^I \wedge \delta_2 e_{I} \right)
+d\delta_{1}e^I \wedge\delta_{2}e_{I}-A_{IJ}\wedge e^I\wedge\delta_2e^J\nn
&=&-d\left( \delta_1 e^I \wedge \delta_2 e_{I} \right)+\delta_{2}e_{I}\wedge
\left( d\delta_{1}e^I +A_{IJ}\wedge \delta_1 e^J\right) \nn &=&-d\left(
\delta_1 e^I \wedge \delta_2 e_{I} \right)+\delta_{2}e_{I}\wedge \left(
e^J\wedge \delta_1 A_{IJ} \right).
\end{eqnarray}
where in the third line in the above equation, we have used the equation of
motion, \emph{i.e.}, the $\delta_1$ version of  eqn. (\ref{HP2}).  Thus, the
term in the L.H.S. is equal to the term on the R.H. S. In other words, 
the Palatini and the Holst
symplectic structure are equivalent when there is no boundary in the spacetime.

If the spacetime has an inner boundary which is the present case of interest,
it is instructive to check wheather the equivalence still holds.
To check that, we go back to the construction of the symplectic current
$J(\delta_{1}, \delta_{2})$. The result has been  calculated before and gives
\begin{equation} J\left( \delta_1, \delta_2
\right):=\dfrac{1}{16\pi G\gamma}\left\lbrace \delta_{[1}\left( e_{1}\wedge
e_{2}\right) \right\rbrace \wedge\left\lbrace \delta_{2]}\left(
A_{IJ}-\frac{\gamma}{2} \epsilon_{IJ}{}^{KL}A_{KL}\right)\right\rbrace
\end{equation}
Integrating the symplectic current over $\mathcal{M}$, taking the orientation
into account, we get:
\begin{equation}
(\int_{M_{+}}-\int_{M_{-}})~J(\delta_{1}, \delta_{2})+ \int_{\Delta}~J(\delta_{1}, \delta_{2})=0
\end{equation}

We have already seen that the $\gamma$- dependent term in the symplectic current
gives a total derivative term. In the following steps, we will only concern
ourselves with $\gamma$- dependent term in the symplectic current since the
other $\gamma$- independent term is the standard Palatini symplectic current.

The first integration of the $\gamma$- dependent term over $M_{+}$ will
go to the boundaries of $M_{+}$ and thence leave an integral over $S_{+}$
(and at infinity which goes to  zero by asymptotic boundary conditions).
Similarly, the integration of the $\gamma$- dependent term over $M_{-}$ will 
leave an integral over $S_{-}$. The integral over $\Delta$ will give
two boundary integrals, one at $S_{+}$ and another at $S_{-}$. Taking the
orientations of the surfaces, we get that the $\gamma$- dependent symplectic current
with $J_{\gamma}(\delta_{1}, \delta_{2})=d\alpha(\delta_{1}, \delta_{2})$ gives:
\begin{equation}
(\int_{S_{+}}-\int_{S_{-}})~\alpha(\delta_{1}, \delta_{2}) + (\int_{S_{-}}-\int_{S_{+}})~\alpha(\delta_{1}, \delta_{2})
+ \int_{M_{+}\cup M_{-}\cup\Delta}J_{P}(\delta_{1}, \delta_{2})=0
\end{equation}
where, $J_{P}(\delta_{1}, \delta_{2})$ is the symplectic current for the Palatini action. 
The above equation shows that the contributions from the $\gamma$- dependent terms
cancel. This implies that even in presence of boundaries, the Holst
symplectic current is equivalent to that of Palatini. In other words,
even in presence of boundaries, the canonical transformation 
holds good.
\subsection{Calculation of $dV$}
\label{sec:calc_dv}
Let us consider the definition of the Riemann tensor:
\begin{equation}
 \left[ \nabla_{a}\nabla_{b}-\nabla_{b}\nabla_{a} \right] X^{c}=-R_{abd}^{c}X_{d}
\end{equation} 

Now, consider the case when the vector $X^a=m^a$. Then, we have:
\begin{eqnarray}
 \left[ \nabla_{a}\nabla_{b}-\nabla_{b}\nabla_{a} \right] m^{c}&\=&\left( \nabla_{a}U_{b}
-\nabla_{b}U_{a} \right)\l^c +\left( \nabla_{a}V^{(m)}_{b}
-\nabla_{b}V^{(m)}_{a} \right)m^c +\left( \omega_{a}U_{b}
-\omega_{b}U_{a} \right)\l^c+ \left(U _{a}V^{(m)}_{b}
-U_{b}V^{(m)}_{a} \right)\l^c\nn
&\=&-R_{abd}{}^{c}~m_{d}
\end{eqnarray} 
Multiply both sides by $\bm_{c}$, we get:
\begin{equation}
 \left( \nabla_{a}V^{(m)}_{b}-\nabla_{b}V^{(m)}_{a} \right)\=-R_{abd}{}^{c}~m^{d}\bm_{c}\=R_{abcd}~m^{d}\bm^{c}
\end{equation} 
Now, we use the expansion of Riemann tensor in terms of the Ricci, the Weyl  tensor and the Ricci Scalar 
\begin{equation}
R_{ab}{}^{cd}=C_{ab}{}^{cd}+2R_{[a}{}^{[c}g_{b]}{}^{d]}-\frac{1}{3}R g_{[a}{}^{c}g_{b]}{}^{d}
\end{equation} 
Then, using the expansion of the the Weyl tensor in terms of the Newman-Penrose scalars (see appendix (\ref{appb})),
we get the following result:
\begin{equation}\label{exp_real_psi} 
 R_{\ub{ab}cd}m^d \bm^c\=4 \mathrm{Re}\mathbf{\Psi_2}m_{[a}\bm_{b]}+\frac{1}{2}\left( R_{bd}\bm_{a} m^{d}- R_{bc}\bm^cm_{a}-R_{da} m^d \bm_{b}+R_{ca}\bm^c m_{b}  \right) +\frac{1}{3} R m_{[a}\bm_{b]}
\end{equation} 
To simplify the second term in the  term in the above expression of eqn. (\ref{exp_real_psi}) , we consider the following. Let,
\begin{equation}
R_{\ub{b}}{}^{d}m_{d}\=Am_{b}+B\bm_{b}+Cn_{b}.
\end{equation} 
where, $A, B$ and $C$ are to be determined. This implies that $A=R_{bd}\bm^{b}m^d$, 
$B=R_{bd}m^{b}m^d$ and $C=0$ as $R_{ab}\l^a m^{b}\=0$. This means,
\begin{equation}
 R_{\ub{b}}{}^{d}m_{d}\=(R_{pq}\bm^{p}m^q) m_{b}+(R_{pq}m^{p}m^q) \bm_{b}
\end{equation} 
Putting these expressions in the eqn. (\ref{exp_real_psi}), we get
\begin{equation}
 dV^{(m)}\=\frac{1}{i}\left[2\mathrm{Re} \mathbf{\Psi_2} -(R_{pq}m^{p}\bm^q)+\frac{R}{6}\right]{}^2\mbf\epsilon 
\end{equation} 
Using the expression $\Phi_{11}=\frac{1}{4}R_{ab}~(\l^{a}n^{b}+m^{a}\bar m^{b})$,
it is easy to check that $\Phi_{11}=\frac{1}{8}R + \frac{1}{2} R_{ab}\l^{a}n^{b}$. Then, defining the real connection $iV^{(m)}=:\bar V^{(m)}=-i m_{[I}\bar m_{J]}A{}^{IJ}$, we get the following
expression:
\begin{equation}
d\bar V^{(m)}\=2(\mathrm{Re}\mathbf{\Psi_2} -\Phi_{11}-\frac{R}{24}){}^2\mbf\epsilon  
\end{equation}
Let us now look at the expression
for the connection $A_{IJ}^{H}$. We shall define a 
new connection for this total connection by projecting $A_{IJ}^{H}$ as follows.
Define the following connection $V^{(H)}\=-i m_{[I}\bar m_{J]}A^{(H)}{}^{IJ}$ and
the curvature of this connection becomes\footnote{See the appendix \ref{appendix_canonical_contact} for details and other comparisons.}:
\begin{eqnarray}\label{exp_dv_epsilon}
dV^{(H)}&=&~\left[ \mathrm{Re}\mathbf{\Psi_2}+ \gamma\mathrm{Im}\mathbf{\Psi_2} -\Phi_{11}-\frac{R}{24}\right]{}^2\mbf\epsilon \nn
&=&~\left[{\mathbf{\Psi_{2}}}^{(H)} -\Phi_{11} - \frac{R}{24}\right]{}^2\mbf{\epsilon} . 
\end{eqnarray} 
where, we have defined ${\mathbf{\Psi_{2}}}^{(H)}= \mathrm{Re}\mathbf{\Psi_2}+ \gamma\mathrm{Im}\mathbf{\Psi_2}$.
If we are in the vacuum $(\Phi_{11})=0$ and the cosmological constant is zero, then we have:
\begin{equation}
 dV^{(H)}\= \mathbf{\Psi_2}^{(H)}~{}^2\mbf{\epsilon} 
\end{equation} 
%

\subsection{Going back to Canonical Phase Space}\label{appendix_canonical_contact}
We now use the foregoing results of the covariant phase space to get the
results of the canonical phase space to make comparisons. To set the stage, we recapitulate
the following conventions and results \cite{Ashtekar_Lewandowski}:

The indices $i,j,\cdots$ take values of the subspace orthogonal
to the internal fixed vector $\tau^I$. If the projection vector is denoted 
by $q^{i}_{I}$, then the induced internal metric
on the subspace is given by:
\begin{equation}
\eta_{ij}=q_{i}^{I}q_{j}^{J}\eta_{IJ}.
\end{equation}
The internal $4-$ dimensional antisymmetric tensor $\epsilon_{IJKL}$ naturally
induces the completely antisymmetric
tensor on the subspace and will be denoted by
$\epsilon_{ijk}$  such that:
\begin{equation}
\epsilon_{ijk}=q_{i}^{I}q_{j}^{J}q_{k}^{K}\tau^{L}\epsilon_{LIJK}.
\end{equation}

If we define the connection $A^{(H)}{}_{IJ}$, then we can define
a connection one form $\Gamma^{i}:=\frac{1}{2}q_{I}^{i}\epsilon^{IJ}{}_{KL}
\tau_{J}A^{(H)}{}^{KL}$ and the extrinsic curvature $K^{1}:=q^{i}_{I}A^{(H)}{}^{IJ}n_{J}$.
Let us now define the connection $A^{(H)}{}^{i}:= \Gamma^{i}- \gamma K^{i}$.
It then follows that 
\begin{equation}
A^{(H)}{}^{i}=-\frac{1}{2}\epsilon^{i}{}_{KL}A^{(H)}{}^{KL}.
\end{equation}

To see further, let us now consider the way to define the unit time- like
normal $\tau^{I}$. The obvious way is to use the null- normals $\l^{a}$
and $n^a$. We define:
\begin{equation}
\tau^{I}:=\frac{(\l^{I}+n^{I})}{\sqrt 2} ~~~ r^{I}:=\frac{(\l^{I}-n^{I})}{\sqrt 2}
\end{equation}

This implies that the internal vector $r^{i}$ is such that it picks out the
connection intrinsic to the two sphere $S_{\Delta}$:
\begin{eqnarray}
A^{i}r_{i}&=&-\frac{1}{2\sqrt 2}q^{I}_{i}(\l_{I}-n_{I})\epsilon^{i}_{KL}A^{(H)}{}^{KL} \\ \nonumber
&=&\frac{i}{4}(\l_{K}n_{L}-\l_{L}n_{K})\epsilon^{KL}{}_{IJ}A^{(H)}{}^{IJ}\\ \nonumber
A^{i}r_{i}&=&-i m_{[I}\bar m_{J]}A^{(H)}{}^{IJ}=:\tilde{V}.
\end{eqnarray}
We can also define momentum $P^{a}_{i}=\frac{1}{16\pi G\gamma}e^{j}_{b}e^{k}_{c}\epsilon_{ijk}\eta^{abc}$. The $2$ form dual to the momentum pulled back to
$S_{\Delta}$ is given by $\ub{\Sigma}^{i}_{ab}=\eta_{abc}P^{a}_{j}\eta^{ij}$.
Then, we have:
\begin{eqnarray}
\Sigma^{i}_{pq}r_{i}&=&\frac{1}{16\pi G\gamma}2im^{[j}\bar m^{k]} \epsilon^{i}_{jk}r_{i}~{}^{2}\epsilon_{pq}\\ \nonumber
&=&\frac{1}{8\pi G\gamma}~{}^{2}\epsilon_{pq}.
\end{eqnarray}

This leads to the following form of the curvature on the sphere $S_{\Delta}$
\begin{equation}\label{epsilon_exp_sphere1}
~dV^{(H)}_{s}=-\frac{2\pi}{\mathcal{A}^{s}_{\Delta}}8\pi G \gamma~ \Sigma^{i}r_{i}
\end{equation} 
%

\subsection{Spherical Symmetry}
We consider the spherical symmetric metric and prove
the claims that $\lambda=\pi\=0$ and $\mu$ is spherically symmetric.

Consider the following most general spherically symmetric
metric:
\begin{equation}
ds^{2}=-f(r,t)~dt^2 + g(r,t)dr^2 +r_{\Delta}^2 d\Omega_{2}
\end{equation}
where, $d\Omega_{2}$ is the $2-$ sphere metric. The null normals can be
calculated
to as follows:
\begin{eqnarray}
\l^{a}&\= &\frac{1}{\sqrt 2}(-\frac{1}{f}~\frac{\partial}{\partial
t}+\frac{1}{g}~\frac{\partial}{\partial r})\nn
n^{a}&\= &\frac{1}{\sqrt 2}(-\frac{1}{f}~\frac{\partial}{\partial
t}-\frac{1}{g}~\frac{\partial}{\partial r})\nn
m^{a}&\= &\frac{1}{r_{\Delta}\sqrt 2}(\frac{\partial}{\partial
\theta}-\frac{i}{Sin\theta}~\frac{\partial}{\partial \phi})
\end{eqnarray}
where, the normalizations have been fixed so that $\l.n=-1$ and $m.\bm=1$. The covariant derivative of the null normal turns out to be:
\begin{equation}
\nabla_{a}n_{b}=\partial_{a}f~(dt)_{b}-f~\Gamma_{ab}^{t}+\partial_{a}g~(dr)_{b}-g~\Gamma_{ab}^{r}
\end{equation}
Then, it follows simply that $\pi=0$ and $\lambda=0$. For $\mu$, observe that
$\mu=(f~\Gamma_{ab}^{t}+g~\Gamma_{ab}^{r})m^{a}\bm^{b}$ will only contribute
for $a,b=\theta,\phi$. Then, $\mu=\frac{2}{r~g}$. Since $g$ is only a
function of
$(t,r)$, $\mu$ is spherically symmetric.


\end{document}